\documentclass[12pt,a4paper,twoside,nopreprintline]{elsarticle}
\usepackage{tkz-euclide}
\usepackage{amsmath,amssymb,amsthm}
\usepackage{hyperref}
\usepackage[strings]{underscore}
\usepackage{esint}
\usepackage[T1]{fontenc}
\usepackage{geometry}

\makeatletter
\def\@oddhead{\mbox{\emph{Solution of all quartic  matrix models}}}
\def\@evenhead{\mbox{\sc H. Grosse, A. Hock and R. Wulkenhaar}}
\makeatother

\title{Solution of all quartic  matrix models}

\author[1,2]{Harald Grosse}
\ead{harald.grosse@univie.ac.at}

\author[3]{Alexander Hock}
\ead{alexander.hock@unige.ch}

\author[4]{Raimar Wulkenhaar}
\ead{raimar@math.uni-muenster.de}

\affiliation[1]{organization={Fakult\"at f\"ur Physik, Universit\"at Wien},
  addressline={Boltzmanngasse 5},
  postcode={A-1090},
  city={Vienna},
  country={Austria}}

\affiliation[2]{organization={Erwin Schr\"odinger International Institute 
for Mathematics and Physics, University of Vienna},
addressline={Boltzmanngasse 9}, postcode={A-1090},  city={Vienna},
country={Austria}}

\affiliation[3]{organization={Section of Mathematics, University of Geneva},
addressline={Rue du Conseil-G\'{e}n\'{e}ral 7-9},
postcode={1205},
\city={Geneva},
\country={Switzerland}}

\affiliation[4]{organization={Mathematisches~Institut,
    Universit\"at~M\"unster}, 
  addressline={Einsteinstr.~62},
  postcode={D-48149},
  city={M\"unster},
  country={Germany}}

\theoremstyle{plain}
 \newtheorem{theorem}{Theorem}[section]
\newtheorem{lemma}[theorem]{Lemma}
\newtheorem{corollary}[theorem]{Corollary}
\newtheorem{proposition}[theorem]{Proposition}
\newtheorem{definition}[theorem]{Definition}

\theoremstyle{definition} 
 \newtheorem{remark}[theorem]{Remark}

\DeclareMathOperator{\Res}{Res}

\numberwithin{equation}{section}

\allowdisplaybreaks[3]
\begin{document}
\newgeometry{twoside,
  paperwidth=210mm,
  paperheight=297mm,
  textheight=23cm,
  textwidth=16cm,
  centering,
  headheight=50pt,
  headsep=1cm,
  footskip=1cm,
  footnotesep=24pt plus 2pt minus 12pt,
 }

\begin{abstract}
  We consider the quartic analogue of the Kontsevich model, which is
  defined by a measure
  $\exp(-{N}\,\mathrm{Tr}(E\Phi^2+(\lambda/4)\Phi^4)) d\Phi$ on
  Hermitian ${N}\times{N}$-matrices, where $E$ is any positive matrix
  and $\lambda$ a scalar.  It was previously established that the
  large-$N$ limit of the second moment (the planar two-point function)
  satisfies a non-linear integral equation. By employing tools from
  complex analysis, in particular the Lagrange-B\"urmann inversion
  formula, we identify the exact solution of this non-linear problem,
  both for finite $N$ and for a large-${N}$ limit to unbounded
  operators $E$ of spectral dimension $\leq 4$. For finite $N$, the
  two-point function is a rational function evaluated at the preimages
  of another rational function $R$ constructed from the spectrum of
  $E$. Subsequent work has constructed from this formula a family
  $\omega_{g,n}$ of meromorphic differentials which obey
  blobbed topological recursion.  For unbounded operators $E$,
  the renormalised two-point function is given by an integral formula
  involving a regularisation of $R$. This allowed a proof, in subsequent work,
  that the $\lambda\Phi^4_4$-model on  noncommutative Moyal space
  does not have a triviality problem.
\end{abstract}


\maketitle

\section{Introduction}

For a positive $N\times N$-matrix $E=\mathrm{diag}(E_1,...,E_N)$, consider
the
Gau\ss{}ian probability measure 
\begin{align}
  d\mu_E(\Phi):=\frac{\exp(-N\,\mathrm{Tr}(E\Phi^2)) d\Phi}{
\displaystyle \int_{H_N}     \exp(-N\,\mathrm{Tr}(E\Phi^2)) d\Phi}
\end{align}
on the space $H_N$ of self-adjoint $N\times N$-matrices,
where $d\Phi$ is the Lebesgue measure on $H_N$. Then
\begin{align}
  \mathcal{Z}_{E,\frac{\mathrm{i}}{3}\Phi^3}
    =\int_{H_N} \!\!\! d\mu_E(\Phi)\, \exp\Big(\frac{\mathrm{i}N}{3}\,
  \mathrm{Tr}(\Phi^3)\Big)
\end{align}
is the generating function of ribbon graphs with
3-valent vertices in which an edge that separates faces with labels $i,j
\in \{1,...,N\}$ carries the weight $\frac{1}{E_j+E_j}$, with summation over
face labels. It was proved by Kontsevich
\cite{Kontsevich:1992ti} 
that 
$\mathcal{Z}_{E,\frac{\mathrm{i}}{3}\Phi^3}$ is, in fact, a function
of `time variables'
$t_k=-(2k-1)!! \mathrm{Tr}(E^{-2k-1})$, and in these
time variables the generating function of intersection numbers of
tautological characteristic classes on the moduli space
$\overline{\mathcal{M}}_{g,n}$ of stable complex curves.
Kontsevich also proved that $\mathcal{Z}_{E,\frac{\mathrm{i}}{3}\Phi^3}$,
as function of
$\{t_k\}$, is
a $\tau$-function of the KdV integrable hierarchy, thus proving a
famous conjecture \cite{Witten:1990hr} due to Witten.

More generally, one can consider moments of diagonal matrix entries
\begin{align}
  \mathcal{M}_{E,\frac{\mathrm{i}}{3}\Phi^3}(k_1,...,k_n)
  = \frac{1}{\mathcal{Z}_{E,\frac{\mathrm{i}}{3}\Phi^3}} 
  \int_{H_N} \!\!\! d\mu_E(\Phi)\; \Phi_{k_1k_1} \cdots
  \Phi_{k_nk_n} 
  \exp\Big(\frac{\mathrm{i} N}{3}\,
  \mathrm{Tr}(\Phi^3)\Big)
  \label{moments-3}
\end{align}
and resulting cumulants. The $1/N$-expansion of
these cumulants can be computed by topological recursion
\cite{Eynard:2007kz, Eynard:2016yaa} from a
spectral curve that is a deformation of the Airy curve $(x=z^2,y=z)$.

Note that $d\mu_E\exp(\frac{\mathrm{i}N}{3}\mathrm{Tr}(\Phi^3))$ is
only a signed measure. Changing it into 
$d\mu_E\exp(\pm \frac{N}{3}\mathrm{Tr}(\Phi^3))$ is
not an option because the corresponding integrals do not converge.
It would therefore be desirable to extend structures established for
the moments (\ref{moments-3}) to
\begin{align}
  \mathcal{M}_{E,P(\Phi)}(k_1l_1,...,k_nl_n)
  &= 
  \frac{\displaystyle \int_{H_N} \!\!\! d\mu_E(\Phi) \;\Phi_{k_1l_1} \cdots
  \Phi_{k_nl_n} 
  \exp(-N\,\mathrm{Tr}(P(\Phi))) }{
\displaystyle
  \int_{H_N} \!\!\! d\mu_E(\Phi) \,
  \exp(-N\,\mathrm{Tr}(P(\Phi)))}
\;,
  \label{moments-P}
\end{align}
where $P$ is a polynomial of \emph{even} degree, real coefficients and
positive coefficient of the top degree. The simplest case is
$P(\phi)=\frac{\lambda}{4}\Phi^4$.  A large zoo of matrix
models has been studied since the 1990s (we refer to 
\cite{DiFrancesco:1993cyw} for an overview about the first period).
Nevertheless, the desirable class (\ref{moments-P}) is
missing so far\footnote{The Kontsevich model can be transformed
  into a matrix model
  with external field. In the class of external field matrix models
  there is also a generalisation of the Kontsevich model to quartic
  (or any other) potential, but this is \emph{not} related to the
  matrix model studied here. See the discussion in sec.~2.1 of
\cite{Branahl:2020uxs}.}.  The reason is that this case is
surprisingly difficult and different.

In this paper we establish the entrance into matrix models 
(\ref{moments-P}) for the simplest case
$P(\Phi)=\frac{\lambda}{4}\Phi^4$:
\begin{theorem}
  \label{thm:intro}
  Let $e_1,...,e_d$ be the pairwise different eigenvalues of $E$
  and $P(\Phi)=\frac{\lambda}{4}\Phi^4$.
There is a ramified covering $R:\mathbb{P}^1\to \mathbb{P}^1$
of degree $d+1$ such that the $1/N$-leading part
\[
  G^{(0)}_{ij}= \frac{1}{N} \mathcal{M}_{E,\frac{\lambda}{4}\Phi^4}(ij,ji)+
  \mathcal{O}(N^{-2})
\]
of the
second moment
is an explicitly given rational function in the preimages
$\{\widehat{\varepsilon_n}^l\}_{n=1,..,d,\;l=0,...,d}$
of the $\{e_n\}$ under $R$, i.e.\ solutions of
$R(\widehat{\varepsilon_n}^l)=e_n$.

\end{theorem}

Because of a recursive structure which is typical for 
matrix models, the formal $1/N$-expansion of any other moment/cumulant
of the quartic matrix model can be obtained from 
$G^{(0)}_{ij}$ by solving \emph{affine} equations \cite{Grosse:2012uv}.
To implement this in practice, some auxiliary functions
$\Omega^{(g)}_{k_1,...,k_n}$ are necessary \cite{Branahl:2020yru},
and precisely those relate
to (a variant of) topological recursion. Any planar cumulant is a
sum of fractions, encoded in nested Catalan tables \cite{deJong:2019oez},
with 2-point functions
$ G^{(0)}_{ij}$ in the numerator and differences $e_k-e_l$ in the denominator.

In fact we prove a far more general result for what we call \emph{all}
quartic matrix models. Recall that in the case of the Kontsevich
model \cite{Kontsevich:1992ti}, the explicit solution of the moment
$\mathcal{M}_{E,\frac{\mathrm{i}}{3}\Phi^3}(k_1)$ was initially found
in \cite{Makeenko:1991ec} in a different context.  Makeenko and
Semenoff replaced the loop equation for
$\mathcal{M}_{E,\frac{\mathrm{i}}{3}\Phi^3}(k_1)$ by a non-linear
integral equation for a sectionally holomorphic function and solved
the resulting Riemann-Hilbert problem by boundary value
techniques. With the Makeenko-Semenoff result
\cite{Makeenko:1991ec} at disposal one can
easily write down an ansatz by which the equation for
$\mathcal{M}_{E,\frac{\mathrm{i}}{3}\Phi^3}(k_1)$ is solved directly.
Otherwise the right ansatz is by no means obvious.

The same strategy worked in the quartic model. We start in this paper
from a more general non-linear integral equation (\ref{eq:Gint}),
established in \cite{Grosse:2009pa} and further analysed in
\cite{Grosse:2012uv}, which in case of Dirac measures reduces to the
loop equation for $G^{(0)}_{ij}$. We introduce in Def.~\ref{def:R} a
class of (sectionally) holomorphic functions $R$, strongly related to
almost general Herglotz-Nevanlinna functions. We show how to use
complex analysis and Lagrange-B\"urmann inversion to evaluate certain
integrals involving such Herglotz-Nevanlinna functions. Then, under a
H\"older condition for the measure, a particular integral $\Psi$,
given in (\ref{def:Psi}) in terms of $R$ and its inverse $R^{-1}$, has
boundary values (\ref{G-formula}) which precisely satisfy the
non-linear integral equation we are interested in. Only some matching
of parameters is necessary (which requires some thought, the
`renormalisation', when extension to half-infinite support is
desired). As result, we establish a one-to-one correspondence between
Herglotz-Nevanlinna functions in which the measure has support in
$[M^2,\infty)$, with $M>0$, and quartic matrix models. In the case of
Dirac measures one can evaluate $\Psi$ by the residue theorem and
finds the result described in Theorem \ref{thm:intro}.

This knowledge permits,
a posteriori, an ansatz that leads to a rather elementary solution
\cite{Schurmann:2019mzu} of the loop equation for $G^{(0)}_{ij}$. But
without the prior work (in the preprint) of the present paper, the
investigation \cite{Schurmann:2019mzu} would have been impossible. It
was subsequently understood \cite{Branahl:2020yru} that the 
ramified covering $R$ plays the r\^ole of the function $x$ of
topological recursion, and that the other function $y$ of the spectral
curve \cite{Eynard:2007kz} is related to $\sum_i
G^{(0)}_{ij}$. However, the recursive structure of the $1/N$-expansion
of the quartic matrix model is not exactly given by topological
recursion. As shown in \cite{Branahl:2020yru} one needs to work within
the more general blobbed topological recursion due to Borot and
Shadrin \cite{Borot:2015hna}.

The more general solution established in this paper is decisive for
quantum field theory on noncommutative geometries. To treat the
divergences in such a QFT, a regularisation to a matrix model is
necessary ---  in an intermediate step. In the end the limit back to
operators on Hilbert space must be taken. This limit destroys the
algebraic structures of matrix models: isolated poles and ramification
points accumulate to branch cuts. The more general approach via
boundary value techniques, employed here and in
\cite{Makeenko:1991ec}, is the only viable road.  In our subsequent
work \cite{Grosse:2019qps} we identified the function $R$ for the
$\lambda\Phi^4$-QFT model on 4-dimensional noncommutative Moyal space.
It is given by a Gau\ss{} hypergeometric function which has, \emph{and this
exceptional for a 4-dimensional model}, a global inverse $R^{-1}$
on $\mathbb{R}_+$. Therefore, the second moment of the
$\lambda\Phi^4_4$-measure is globally defined by our explicit formula
(\ref{G-formula}) for any coupling constant $\lambda>-\frac{1}{\pi}$.
This is in sharp constrast to the standard $\lambda\phi^4_4$-model which
is marginally trivial \cite{Aizenman:2019yuo} and as such impossible to
construct. Even better, the effective spectral dimension is 
 reduced from the na\"ive value $4$ to
$4-\frac{2}{\pi}\arcsin(\lambda \pi)$. It would be interesting
to investigate whether the reduced spectral dimension, consequence of our
exact solution of the two-point function, admits to transfer the spectacular
methods and results \cite{Hairer2014, Mourrat:2016vlt, Gubinelli:2021nou}
of the ordinary $\lambda\phi^4_3$-model to
the 4-dimensional noncommutative case.

\subsection*{Organisation of the paper}

In sec.~\ref{sec:setup} we recall from \cite{Grosse:2009pa, Grosse:2012uv} the
non-linear equation for the planar two-point function $G$. To solve it
we introduce in sec.~\ref{sec:R} an auxiliary function $R$ and evaluate
in sec.~\ref{sec:R-integral} via Lagrange-B\"urmann inversion several
integrals involving $R$. We show in sec.~\ref{sec:G} that boundary
values of these integrals provide the solution $G$ (as integral
formula involving $R$ and its inverse $R^{-1}$) of the given non-linear
equation. The renormalisation procedure is described in
sec.~\ref{sec:renormalisation}. In sec.~\ref{sec:finiteN} we specify to finite
matrices and show that the integral formula for $G$ can be evaluated
explicitly. The rationality result of Theorem~\ref{thm:intro} refers to
(\ref{G-partialfraction}), but also the equivalent
representation (\ref{Gzw-rational}) is of interest.
A few examples are given in sec.~\ref{sec:ex}.
We finish by a longer epilogue  (sec.~\ref{sec:out}) which
puts the result of this paper in relation to the quest for
interacting quantum field theories and gives an outlook to subsequent work
related to blobbed topological recursion.

\subsection*{Acknowledgements}
       
RW would like to thank Erik Panzer for the joint solution of a special
case which is indispensable prerequisite of the present paper. AH
thanks Akifumi Sako for hospitality during a visit of the Tokyo
University of Science where first thoughts to generalise the special case
were developed. AH also thanks the University of Oxford for providing
an excellent research envirenment during a Walter Benjamin fellowship.
This work was supported by the Erwin
Schr\"odinger Institute \mbox{(Vienna)} through a ``Research in Team''
grant and by the Deutsche Forschungsgemeinschaft via the Cluster of
Excellence\footnote{``Gef\"ordert durch die Deutsche
  Forschungsgemeinschaft (DFG) im Rahmen der Exzellenz\-strategie des
  Bundes und der L\"ander EXC 2044 –390685587, Mathematik M\"unster:
  Dynamik--Geometrie--Struktur''} ``Mathematics M\"unster'' and the RTG
2149.

\section{The setup}
\label{sec:setup}

Let $E$ be a positive (as operator on Hilbert space) $N\times N$-matrix
and $\lambda >0$ a scalar. We consider the second
moment $ZG_{ab}$ of the quartic matrix model
with Kontsevich-type covariance,
\begin{align}
ZG_{ab} := \frac{1}{{N}}
\frac{\displaystyle 
\int_{H_N}\!\!\!\!\!
d\Phi \; 
\Phi_{ab} \Phi_{ba}
\exp\Big({-}{N} \mathrm{Tr}\big(E\Phi^2  + \tfrac{\lambda 
}{4}\Phi^4 \big)\Big)}{\displaystyle 
\int_{H_N}\!\!\!\!\!
d\Phi \; 
\exp\Big({-}{N} \mathrm{Tr}\big(E\Phi^2  + \tfrac{\lambda 
}{4}
\Phi^4 \big)\Big)}\,.
\label{eq:ZGab}
\end{align}
The r\^ole of $Z$ and $\mu_{bare}$ (introduced soon)
will be explained below; for fixed $N$ one can set $Z=1$ and
$\mu_{bare}=0$.
The second moment has a 
formal $1/N$ expansion $G_{ab}=\sum_{g=0}^\infty N^{-2g}
G_{ab}^{(g)}$ (which is typical for matrix models).
It was proved in \cite{Grosse:2009pa,Grosse:2012uv} that
the leading contribution
$G_{ab}^{(0)}$, the \emph{planar 2--point function},
satisfies the closed equation
\begin{align}
ZG_{ab}^{(0)}&=\frac{1}{E_a+E_b}-\frac{\lambda}{{N}(E_a+E_b)}
\sum_{n=1}^{{N}} 
\Big( ZG_{ab}^{(0)} \;ZG^{(0)}_{an}
- \frac{ZG_{nb}^{(0)} -ZG_{ab}^{(0)} }{E_{n}-E_a}\Big)\;.
\label{Gab-orig}
\end{align}
Here, $\{E_a\}$ are the eigenvalues of $E$, and the $\Phi_{ab}$ in
(\ref{eq:ZGab}) are the matrix elements of $\Phi$ in the eigenbasis of $E$.
A pedestrian derivation of (\ref{Gab-orig}) is given
in appendix A of \cite{Schurmann:2019mzu}.
The key observation is that, writing
\begin{align}
G_{ab}^{(0)}:= G(x,y)\Big|_{x=E_a-\mu^2_{bare}/2,\;y=E_b-\mu^2_{bare}/2}\;,
\label{Gab-continuation}
\end{align}
then $G(x,y)$ originally defined only on the (shifted) spectrum of $E$
extends\footnote{Such an extensions are instrumental to relate to
  spectal curces in topological recursion \cite{Eynard:2007kz,
    Eynard:2016yaa}.  For the present case, this extension is
  discussed in detail in sec.~3.1 of \cite{Branahl:2020yru}.}  to a
sectionally holomorphic function which satisfies the non-linear
integral equation
\begin{align}
&(\mu_{bare}^2{+}x{+}y)ZG(x,y)
\label{eq:Gint}
\\
&= 1-\lambda
\int_{\tilde{M}^2}^{\tilde{\Lambda}^2} \!\! dt\;\rho_0(t)
\Big( ZG(x,y) \;ZG(x,t) 
- \frac{ZG(t,y) -ZG(x,y)}{t-x}\Big) \;.
\nonumber
\end{align}
The interval $[\tilde{M}^2,\tilde{\Lambda}^2]$ is chosen such that
it contains $\{E_n-\frac{\mu_{bare}^2}{2}\}$, and we have defined
\begin{align}
  \label{rhot}
  \rho_0(t)&:=\frac{1}{{N}} \sum_{n=1}^{{N}}
\delta\Big(t-\Big(E_n{-}\frac{\mu^2_{bare}}{2}\Big) \Big)\;.
\end{align}
Following \cite{Panzer:2018tvy} one can also derive 
a symmetric equation equivalent to (\ref{eq:Gint}):
\begin{align}
  \label{eq:Gintsym}
  &(\mu_{bare}^2+x+y)ZG(x,y)
\\
&= 1
+\lambda
\int_{\tilde{M}^2}^{\tilde{\Lambda}^2} \!\!
dt\;\rho_0(t) \,\frac{ZG(t,y) -ZG(x,y)}{t-x}
+\lambda \int_{\tilde{M}^2}^{\tilde{\Lambda}^2} \!\!
ds\;\rho_0(s) \,\frac{ZG(x,s) -ZG(x,y)}{s-y}
\nonumber
\\
&-\lambda^2
\int_{\tilde{M}^2}^{\tilde{\Lambda}^2} \!\! dt\;\rho_0(t) 
\int_{\tilde{M}^2}^{\tilde{\Lambda}^2} \!\! ds\;\rho_0(s) 
\, \frac{ZG(x,y)\,ZG(t,s)
- ZG(x,s)\,ZG(t,y)}{(t-x)(s-y)}\;.
\nonumber
\end{align}

This paper provides the exact solution of the non-linear
equation~(\ref{eq:Gint}).
In fact we solve the problem in a larger quantum field theoretical
perspective. This refers to a limit ${N}\to \infty$ in which
the matrix $E$ becomes an unbounded operator on Hilbert space (consequently,
$E_N\to \infty$ and $\tilde{\Lambda}\to \infty$). For the
Kontsevich model, the same quantum field theoretical extension was
solved in \cite{Grosse:2016pob, Grosse:2016qmk, Grosse:2019nes}. Of
course one can study a large-${N}$ limit in which 
$E-\frac{\mu^2_{bare}}{2}$ is
resized to keep a finite support $[\tilde{M}^2,\tilde{\Lambda}^2]$
of the measure. We
call this the dimension-0 case. It is only little more effort to solve
the problem for two classes (dimension $D=2$ and $D=4$)
of unbounded operators $E$. Our strategy
follows closely the usual renormalisation procedure in quantum field
theory. This means that $\mu_{bare}^2$ and possibly $Z$ are carefully
chosen functions of the data $\{E_N\}$ and $\tilde{\Lambda}$; only
with the right dependence a limit $\lim_{N,E_N,\tilde {\Lambda}\to \infty}G(x,y)$
can be achieved. We give details on the \emph{spectral dimension}
(which captures Weyl's law of the asymptotics of eigenvalues of
the Laplacian) and the precise dependence of
$\mu_{bare}^2, Z$ on  the data $\{E_N\}$ and $\tilde{\Lambda}$ in
sec.~\ref{sec:renormalisation}.

\begin{remark}
Equation (\ref{eq:Gint}) is the analogue of the equation 
\begin{align*}
(W(x))^2 -\lambda^2 \int_0^{\Lambda^2} \!\! dt \;\rho_0(t) \,
\frac{W(t)-W(x)}{t-x}= x\;,\quad
\rho_0(t)=\frac{8}{{N}} \sum_{n=1}^{{N}} \delta(t-(2E_n)^2)
\end{align*}
in the Kontsevich model (in dimension $D=0$; generalised in 
\cite{Grosse:2016pob, Grosse:2016qmk} to $D\in \{2,4,6\}$;
with $\lambda$ the coefficient in the potential
$P(\Phi)=\frac{\mathrm{i}\lambda}{3}\mathrm{Tr}(\Phi)$). 
Its solution found by Makeenko and Semenoff \cite{Makeenko:1991ec} was later 
understood to provide the key ingredients of 
the \emph{spectral curve of topological recursion} 
\cite{Eynard:2007kz, Eynard:2016yaa}. 
The solution is universal in terms of an implicitly defined parameter 
$c$, which depends on $E,\lambda$ and a dimension $D\in \{0,2,4,6\}$
(which we introduce in sec.~\ref{sec:renormalisation}):
\begin{align}
\label{implicit-c}
c 
&= \frac{\lambda^2}{
\big(\frac{2}{1+\sqrt{1+c}}\big)^{\delta_{D,2}+\delta_{D,4}}}
  \int_{\sqrt{1+c}}^\infty \; \frac{\varrho(y)\,dy}{
y(\sqrt{1+c}+y)^{D/2}}\;,\qquad
\\
\varrho(y)&=\frac{8}{{N}} \sum_{n=1}^{{N}}
\delta(y-\sqrt{4E_n^2+c})\;. \nonumber
\end{align}  
This parameter $c$ effectively deforms the initial matrix $E$ to
$\sqrt{E^2+c/4}$ and thereby the measure $\rho_0$ 
into an implicitly defined deformed measure $\varrho$. 
\hspace*{\fill} $\triangleleft$
\end{remark}

We will see that exactly the same is true for the quartic
model. Employing complex analysis
techniques similar to \cite{Makeenko:1991ec}, we prove that 
equations~(\ref{eq:Gint}) or (\ref{eq:Gintsym}) have a 
universal solution in terms of a deformation $\varrho$ of 
the measure $\rho_0$ given in (\ref{rhot}).

\section{Solution via boundary value problem}

\label{sec:R}

We will prove in this section that a solution of the
non-linear integral equation (\ref{eq:Gint}) can be found in terms of
an auxiliary function $R$ introduced in (\ref{def:Rnew}) below.
It seems surprising that the solution
succeeds in this way. We arrived at this strategy in the converse
order than presented here. The reformulation of (\ref{eq:Gint}) as
a boundary value problem and expression in terms of an angle function
was worked out already in \cite{Grosse:2012uv} and \cite{Panzer:2018tvy}.
This angle function appears in
(\ref{tau-R}) below, and the key guess was to make the ansatz
involving $R(y)-R(-x-\mathrm{i}\epsilon)$ for an unknown function $R$.
We then found that in order to solve (\ref{eq:Gint}), this function $R$
must satisfy the identity (\ref{identity-R}). It turns out that
(\ref{def:Rnew}) does the job.

To achieve this we use tools from previous centuries:
\begin{itemize}
\item Lagrange inversion theorem \cite{Lagrange:1770??} and a
  generalisation due to B\"urmann \cite{Buermann:1799??}:
\begin{theorem}
\label{thm:Lagrange-inversion}
Let $\phi(w)$ be analytic at $w=0$ with $\phi(0)\neq 0$ and 
$f(w) := \frac{w}{\phi(w)}$. Then the inverse $g(z)$ of $f(w)$
with $z=f(g(z))$ is analytic at $z=0$ and given by
\begin{equation}
g(z) = \sum_{n=1}^{\infty} \frac{z^n}{n!} 
\frac{d^{n-1}}{d w^{n-1}}\Big|_{w=0} (\phi(w))^n\;.
\label{eq:Lagrange}
\end{equation}
More generally, if $H(z)$ is an arbitrary analytic function 
with $H(0)=0$, then
\begin{equation}
H(g(z)) = \sum_{n=1}^{\infty} \frac{z^n}{n!} 
\frac{d^{n-1}}{d w^{n-1}}\Big|_{w=0} \Big( H'(w) \big(\phi(w)\big)^n \Big)\;.
\label{eq:Buermann}
\end{equation}
\end{theorem}
Historically, these inversion formulae were formulated for
formal power series, but the result is also true for convergent power
series and holomorphic functions.

\item Complex analysis was developed by Cauchy between 1825 and 1831.
The residue theorem was presented by Cauchy in a memoire to the
Academy of Sciences of Turin in 1831. A later reprint can be found
in \cite{Cauchy:1831}. There is no need to recall them.
\end{itemize}

\subsection{Definition}

\begin{definition}
  \label{def:R}
We consider a class of holomorphic functions
  $R:\mathbb{C}\setminus [-\Lambda^2,-M^2]\to
\mathbb{C}$ which admit an integral representation
  \begin{align}
  R(z)=\alpha z+\beta -\lambda \int_{\mathbb{R}} dt\;
  \frac{\varrho(t)}{t+z}\;,
\label{def:Rnew}
\end{align}
where $\varrho$ is a positive finite measure on
  $\mathbb{R}$ with support contained in an interval
  $[M^2,\Lambda^2]$,
  for $0<M<\Lambda$. We require 
$\alpha>0$ and $\beta\in \mathbb{R}$,  and $\lambda \in \mathbb{C}$
to be taken from a neighbourhood of $\mathbb{R}_{\geq 0}$.
\end{definition}

\begin{remark}
For $\lambda>0$, the function $z\mapsto y(z)=-R(-z)$ that will be important
below is the almost general representatation 
\cite{Nevanlinna;1922} of a Herglotz (or Nevanlinna, Pick, R-)
function, i.e.\ a function which is holomorphic on the upper half plane
$\mathbb{H}$ and maps $\mathbb{H}$ to itself. The most general
representation would allow $\varrho$ to have unbounded support on
$\mathbb{R}$, which then requires additional growth conditions on $\varrho$.
The extension to half-infinite support $\Lambda\to \infty$ is precisely the
renormalisation problem discussed in sec.~\ref{sec:renormalisation}.
The integrals we derive in sec.~\ref{sec:R-integral} could be of
interest in the general theory of Herglotz-Nevanlinna functions.
\hfill $\triangleleft$
\end{remark}

  \begin{lemma}
\label{lem:U}
Let $\lambda_- :=-\alpha /\int_{\mathbb{R}} dt \frac{\varrho(t)}{(t-M^2/2)^2}$
(a negative real number).
There is a neighbourbood of $[\lambda_-,\infty)$ such that, for any $\lambda$
in (\ref{def:Rnew}) taken from this neighbourhood, 
$R$ is a biholomorphic map of the right half plane
$H_+:=\{z\;|~\mathrm{Re}(z)>0\}$ to the
domain $\mathcal{V}=R(H_+) \subset \mathbb{C}$.
\begin{proof}
We show that $R$ is injective on $H_+$. 
Any two points $z_0\neq z_1 \in H_+$ can be connected by a straight line 
$[0,1]\ni s \mapsto c(s)=z_0+(z_1-z_0)s \in H_\mu$. 
Then 
\begin{align*}
R(z_1)-R(z_0) &= (z_1-z_0) \Big(\alpha +\lambda \int_0^1 ds 
\int_{\mathbb{R}}
\frac{dt\;\varrho(t)}{(t+ c(s))^2}\Big)
\\
&=(z_1-z_0) \Big(\int_0^1 ds 
\int_{\mathbb{R}}dt\;\varrho(t)
\Big\{\frac{\lambda}{(t+ c(s))^2}
+\frac{|\lambda_-|}{(t-\frac{M^2}{2})^2}\Big\}
\Big)\;.
\end{align*}
If $\lambda\geq \lambda_-$ is real, the part in $\{~\}$ has positive
real part for all $z,z_0$ with $\mathrm{Re}(z),\mathrm{Re}(z_0)\geq 0$.
By continuity, the part in $\{~\}$ keeps a positive real part for $\lambda$
in a neighbourhood of $[\lambda_-,\infty)$. 
A holomorphic and injective map between domains in $\mathbb{C}$
is biholomorphic.
\end{proof}
\end{lemma}
Globally, $\mathbb{H}\ni z\mapsto y(z)=-R(-z)\in \mathbb{H}$ is not 
injective. The corresponding preimages of $R$
will be important in sec.~\ref{sec:finiteN}.

\subsection{Contour integrals}

\label{sec:R-integral}

The following theorem is the main technical step.
\begin{theorem}
  \label{thm:R}
  Let $\Gamma$ be a contour in the complex plane which encircles
  $[M^2,\Lambda^2]$ close enough in clockwise orientation
  (see Figure~\ref{fig:contour}).
Let $\lambda_+>0$ be the parameter for which 
$R(M^2)=\max(0,\beta-\alpha M^2)$, and $\lambda_-<0$ be as in Lemma~\ref{lem:U}.
Then there exists a complex
neighbourhood $\mathcal{L}$
of an open subinterval of $[\lambda_-,\lambda_+]$ that contains $0$ and a
complex neighbourhood $\mathcal{U}$ of $[M^2,\infty)$ such that
for all $\lambda\in \mathcal{L}$ and $z\in \mathcal{U}\setminus [M^2,\Lambda^2]$,
the function $R$ defined in (\ref{def:Rnew}) satisfies
\begin{align}
 \frac{1}{2\pi\mathrm{i}}
\int_{\Gamma} dw\;R'(w)
&\log \big(R(z)- R(-w)\big)
\label{identity-R}
\\[-2ex]
&= 2\beta -R(z)-R(-z)
-\lambda\int_{\mathbb{R}} dt\;
\frac{R'(t) \varrho(t)}{R(t)-R(z)}\;.
\nonumber
\end{align}
\end{theorem}
We prove this theorem in two steps in Lemma~\ref{lem1-forthm} and
Lemma~\ref{lem2-forthm}.
\begin{lemma}
  \label{lem1-forthm}
  The function $R$ given in (\ref{def:Rnew}) satisfies
\begin{align}
\frac{\alpha}{2\pi\mathrm{i}}
\int_{\Gamma} dw\;
\log \big(R(z)-R(-w)\big)  = \alpha z+\beta-R(z)
\label{eq:Kz}
\end{align}
for  all $(\lambda,z)\in \mathcal{L}\times \mathcal{U}$,
where $\Gamma,\mathcal{L},\mathcal{U}$ are the same as in
Thm.~\ref{thm:R}.
\begin{proof}
The proof of Lemma~\ref{lem:U} shows $R'(z)>0$ for all real $z\geq 0$
so that we have $R(z)>\max(0,\beta-\alpha M^2)$ for all
$(\lambda,z)\in (\mathcal{L}\cap \mathbb{R}) \times (\mathcal{U}\cap
\mathbb{R})$. This means
  $R(z)+\alpha w-\beta \notin \mathbb{R}_{\leq 0}$ for all $w\geq M^2$
  and real $(\lambda,z)\in \mathcal{L}\times\mathcal{U}$. By continuity, the neighbourhoods
  $\mathcal{L},\mathcal{U}$ can be chosen such that
  $R(z)+\alpha w-\beta \notin \mathbb{R}_{\leq 0}$ for all
  $z,w\in \mathcal{U}$ and $\lambda\in \mathcal{L}$, and we assume
  such a choice here and for Thm~\ref{thm:R}. Furthermore, we choose
  the contour $\Gamma$ that encircles $[M^2,\Lambda^2]$ inside
  $\mathcal{U}$.  It is depicted on the left of
  Figure~\ref{fig:contour}.
\begin{figure}[ht]
\begin{center}\begin{tikzpicture}
  \tkzDefPoint(-6,0){GO}
  \tkzDefPoint(-6,1){GB}  
  \tkzDefPoint(-5.5,0){GM}
  \tkzDefPoint(-5.5,0.3){GM1}
  \tkzDefPoint(-5.5,-0.3){GM2}
  \tkzDefPoint(-4.8,0){GL}
  \tkzDefPoint(-4.8,0.3){GL1}  
  \tkzDefPoint(-4.8,-0.3){GL2}
  \tkzDrawLine[-Latex, add=1 and 0.5](GO,GB)
  \tkzDrawLine[-Latex, add=0.5 and 0.5](GO,GL)
  \tkzDrawLine[add=0 and 0,line width=2pt](GL,GM)
  \tkzDrawArc[line width=1pt](GL,GL2)(GL1)    
  \tkzDrawArc[line width=1pt](GM,GM1)(GM2)
  \tkzDrawLine[add=0 and 0,line width=1pt,tkz arrow={Latex[scale=0.7]}](GM1,GL1)
  \tkzDrawLine[add=0 and 0,line width=1pt,tkz arrow={Latex[scale=0.7]}](GL2,GM2)
  \tkzDefPoint(0,0){O}
  \tkzDefPoint(0,1){B}  
  \tkzDefPoint(2.1,0){R}
  \tkzDefPoint(0.5,0){M}
  \tkzDefPoint(0.5,0.3){M1}
  \tkzDefPoint(0.5,-0.3){M2}
  \tkzDefPoint(0.5,0.07){M3}
  \tkzDefPoint(0.5,-0.07){M4}  
  \tkzDefPoint(1.2,0){L}
  \tkzDefPoint(1.2,0.3){L1}  
  \tkzDefPoint(1.2,-0.3){L2}
  \tkzDefPoint(1.2,0.07){L3}  
  \tkzDefPoint(1.2,-0.07){L4}
  \tkzInterLC(L3,M3)(O,R)\tkzGetPoints{R1}{R3}
  \tkzInterLC(L4,M4)(O,R)\tkzGetPoints{R4}{R2}  
  \tkzInterLC(L4,M4)(L,L1)\tkzGetPoints{P3}{P2}
  \tkzInterLC(L3,M3)(L,L1)\tkzGetPoints{P1}{P4}
  \tkzInterLC(M2,O)(O,R)\tkzGetPoints{S2}{S1}  
  \tkzFillCircle[gray!20](O,R)
  \tkzFillCircle[white](L,L1)  
  \tkzFillCircle[white](M,M1)
  \tkzFillPolygon[white](L1,M1,M2,L2)
  \tkzFillPolygon[white](P1,R1,R2,P2)      
  \tkzDrawLine[-Latex, add=2.5 and 1.6](O,B)
  \tkzDrawLine[-Latex, add=2.3 and 1.3](O,L)
  \tkzDrawLine[add=0 and 0,line width=2pt](L,M)
  \tkzDrawArc[line width=1pt](L,L2)(P2)    
  \tkzDrawArc[line width=1pt](L,P1)(L1)    
  \tkzDrawArc[line width=1pt](M,M1)(M2)
  \tkzDrawLine[add=0 and 0,line width=1pt,tkz arrow={Latex[scale=0.7]}](M1,L1)
  \tkzDrawLine[add=0 and 0,line width=1pt,tkz arrow={Latex[scale=0.7]}](L2,M2)
  \tkzDrawArc[line width=1pt,tkz arrow={Latex[scale=0.7]}](O,R1)(R2)
  \tkzDrawLine[add=0 and 0,line width=1pt,tkz arrow={Latex[scale=0.7]}](P1,R1)
  \tkzDrawLine[add=0 and 0,line width=1pt,tkz arrow={Latex[scale=0.7]}](R2,P2)
  \tkzDrawLine[->,add=0 and 0](O,S1)
  \tkzLabelLine[above](O,S1){\small$r$}
\end{tikzpicture}\vspace*{-0.5cm}
\end{center}
\caption{Sketch of the integration contours
  $\Gamma$ (left) and
  $\Gamma_r$ (right). 
The fat part of the real axis indicates the 
interval $[M^2,\Lambda^2]$. \label{fig:contour}}
\end{figure}
Thus, $w\mapsto \log(R(z)+\alpha w-\beta)$ is holomorphic in
$\mathcal{U}$ and has vanishing integral over $\Gamma$, for any
$(\lambda,z)\in \mathcal{L}\times \mathcal{U}$. We combine this
vanishing integral with our target and want to prove
\begin{align}
\frac{\alpha}{2\pi\mathrm{i}}
\int_{\Gamma} dw\;
\log \Big(1- \frac{R(-w)+\alpha w-\beta}{
  R(z)+\alpha w-\beta}\Big)
  = \alpha z+\beta-R(z)\;.
\label{eq:Tz}
\end{align}

We extend the contour $\Gamma$ as follows to a contour $\Gamma_r$ (see
the right part of Figure~\ref{fig:contour}).  Let $p$ be the largest
intersection of $\Gamma$ with $\mathbb{R}$.  Starting at $p$, move
along the real axis to $r>p$, follow the circle of radius $r$
counterclockwise back to $r$, then run along $\mathbb{R}$ in negative
direction to $p$ and follow $\Gamma$ in its orientation back to
$p$. The integrals over $[p,r]$ cancel each other because of different
orientation, and the integral over the circle converges to $0$ for
$r\to \infty$ (this was the reason to include the denominator
$R(z)+\alpha w-\beta$). To be precise, one needs to check that
$w\mapsto \log \Big(1- \frac{R(-w)+\alpha w-\beta}{ R(z)+\alpha
  w-\beta}\Big)$ is holomorphic in a neighbourhood of $\Gamma_r$.  The
argument of the logarithm in the integral in (\ref{eq:Tz}) is
\begin{align}
1- \frac{R(-w)+\alpha w-\beta}{
  R(z)+\alpha w-\beta}
= 1 + \frac{\lambda  \int_{\mathbb{R}} dt\;
  \frac{\varrho(t)}{t-w}}{
  \alpha(z+w)+ \lambda  \int_{\mathbb{R}} dt\;
  \frac{\varrho(t)}{t+z}}\;.
\label{argument-log}
\end{align}
For $|w|=r$ and $r$ large enough the real part is positive and
the logarithm well-defined.
For $w=x+\mathrm{i}\epsilon$ and $(\lambda,z)$ real, 
let us write 
$\lambda  \int_{\mathbb{R}} dt\;
  \frac{\varrho(t)}{t-w}=A+B\mathrm{i}$
and $\alpha(z+w)+ \lambda  \int_{\mathbb{R}} dt\;
\frac{\varrho(t)}{t+z}
=C+D\mathrm{i}$.
Then $1+\frac{A+B\mathrm{i}}{C+D\mathrm{i}}$
is real for $A=\frac{BC}{D}$, and at that point
\begin{align}
  1+\frac{A+B\mathrm{i}}{C+D\mathrm{i}}\mapsto 1+\frac{B}{D}
=\frac{1}{\alpha}\Big(\alpha+\lambda 
 \int_{\mathbb{R}} dt\;
 \frac{\varrho(t)}{(t-x)^2+\epsilon^2}\Big)\;.
\label{ABCD}
\end{align}
Note that $(t-x)^2+\epsilon^2$ is larger than the squared distance
between $w\in \Gamma$ and $[M^2,\lambda^2]$. 
For $\lambda\geq 0$ the expression (\ref{ABCD}) is positive and the logarithm
in (\ref{eq:Tz}) well-defined. For $\lambda<0$, the expression
(\ref{ABCD}) stays positive for $\tilde{\lambda}_-<\lambda<0$ for some 
critical value $\tilde{\lambda}_-<0$ that depends on $\Gamma$.
The contour integral (\ref{eq:Tz}), with $\Gamma \mapsto \Gamma_r$, is
then well-defined for real $z>0$ and
real $\lambda>\tilde{\lambda}_-$. By continuity it remains well defined
for $z\in \mathcal{U}$ and $\lambda\in \mathcal{L}$, where
$\mathcal{L}$ is some neighbourhood of an open subinterval of
$[\max(\lambda_-,\tilde{\lambda}_-),\lambda_+]$ that contains $0$.
In this situation,  (\ref{eq:Kz}) is equivalent to
\begin{align}
\frac{\alpha}{2\pi\mathrm{i}}
\int_{\Gamma_r} dw\;
\log \Big(1- \frac{R(-w)+\alpha w-\beta}{
  R(z)+\alpha w-\beta}\Big)
  = \alpha z+\beta-R(z)\;.
\label{eq:R-Gamma-r}
\end{align}

Both sides of (\ref{eq:R-Gamma-r}) are holomorphic
in $\lambda \in \mathcal{L}$. By the identity theorem
of holomorphic functions it is thus enough to prove 
(\ref{eq:R-Gamma-r}) in a small ball $|\lambda|< \lambda_\epsilon$ contained in
$\mathcal{L}$, for some $\lambda_\epsilon>0$.
The argument of the logarithm in (\ref{eq:R-Gamma-r}) is given in
(\ref{argument-log}); it converges to $1$ for $\lambda\to 0$.
We can therefore chose 
$\lambda_\epsilon$ such that 
\[
\Big|\frac{R(-w)+\alpha w-\beta}{
  R(z)+\alpha w-\beta}
\Big|<1
\quad\text{ for all }
\quad |\lambda|<\lambda_\epsilon\;,~
w\in \Gamma_r\;,~z\in \mathcal{U}\;.
\]
For $|\lambda|<\lambda_\epsilon$ we can thus expand the logarithm into
a power series. This series is uniformly convergent on
$ \Gamma_r$ so that integral and series commute.
Denoting the integral in (\ref{eq:R-Gamma-r}) by $K_z$, we have
\begin{align}
K_z&=-\sum_{n=1}^\infty \frac{\alpha}{2\pi\mathrm{i}n}
\int_{\Gamma_r} dw\,
\frac{1}{(R(z)+\alpha w-\beta)^n}
\Big(-\lambda \int_{\mathbb{R}} dt\;
\frac{\varrho(t)}{t-w}\Big)^n \;.
\label{Kz-expand}
\end{align}
We evaluate this integral by the residue theorem.
The integral $\int_{\mathbb{R}} dt\;
\frac{\varrho(t)}{t-w}$ is holomorhic in the interior
of $\Gamma_r$ (shaded in gray in Fig.~~\ref{fig:contour})
so that the only singularity is the $n$-fold pole
at $w = \frac{1}{\alpha}(\beta-R(z))$.
We recall the remark from the beginning of this proof that
$\frac{1}{\alpha}(\beta-R(z))<M^2$ for 
$z\in \mathcal{U}\cap \mathbb{R}$ and
$\lambda\in \mathcal{L}\cap \mathbb{R}$. The pole is is thus located left
of the interval $[M^2,\Lambda^2]$. The contour $\Gamma_r$ can be assumed
to pass between pole and interval. Then, the pole
at $w = \frac{1}{\alpha}(\beta-R(z))$
is located in the interior of $\Gamma_r$ (shaded in gray in
Fig.~~\ref{fig:contour}). The residue theorem
evaluates the integral
(\ref{Kz-expand}) to
\begin{align}
K_z&=-\alpha \sum_{n=1}^\infty \frac{(-\lambda/\alpha)^n}{n!}
\frac{d^{n-1}}{dw^{n-1}}\Big|_{w= 0}  \big(\phi_z(w)\big)^n\;,\qquad\text{where}
\nonumber
\\
\phi_z(w)&:=\int_{\mathbb{R}} dt\;
\frac{\varrho(t)}{t+\frac{1}{\alpha}(R(z)-\beta)-w}\;.
\label{phizw}
\end{align}
The Lagrange inversion formula (\ref{eq:Lagrange}) 
shows that $g_z(-\lambda/\alpha)=-\frac{1}{\alpha}K_z$ is the
inverse solution of the equation
$-\frac{\lambda}{\alpha}= f_z(-K_z/\alpha)$, where $f_z(w)=\frac{w}{\phi_z(w)}$.
This means 
\begin{align}
K_z = \lambda\phi_z(-K_z/\alpha)=
  \lambda \int_{\mathbb{R}} dt\;
\frac{\varrho(t)}{t+\frac{1}{\alpha}(R(z)-\beta+K_z)}\;.
\label{Kz}
\end{align}
Introducing $u(z)=\frac{1}{\alpha}(K_z+R(z)-\beta)$,
equation (\ref{Kz}) reads
\[
R(z)
=\alpha u(z)+\beta
-\lambda \int_{\mathbb{R}} dt\;
\frac{\varrho(t)}{t+u(z)}\;.
\]
The rhs equals $R(u(z))$. For small enough $|\lambda|$,
$u(z)$ stays near $z$, in particular in the half plane $H_+$ where
$R$ is injective. Consequently, 
$u(z)\equiv z$, and (\ref{eq:R-Gamma-r}) is proved. But 
(\ref{eq:R-Gamma-r}) was equivalent to
(\ref{eq:Kz}), and the Lemma is 
proved. 
\end{proof}
\end{lemma}

\begin{lemma}
  \label{lem2-forthm}
  For any integrable function 
  $\varrho$ with support contained in
  $[M^2,\Lambda^2]$ one has
\begin{align}
\frac{1}{2\pi\mathrm{i}}
&\int_{\Gamma} dw\;
\int_{\mathbb{R}} ds \frac{\varrho(s)
}{(s+w)^2}
\log \big(R(z)-R(-w)\big)
\nonumber
\\
&=  \int_{\mathbb{R}} ds\;
\varrho(s)
\Big(\frac{1}{s-z}
- 
\frac{R'(s)}{R(s)-R(z)}\Big)\;,
\label{eq:L}
\end{align}
for  $\lambda\in \mathcal{L}$ and $z\in \mathcal{U}$.
The function $R$ in (\ref{def:Rnew}) depends on the same
function $\varrho$, and $\mathcal{L},\mathcal{U}$ are as in
Theorem~\ref{thm:R}.
\begin{proof}
By the same arguments as in  proof of
Lemma~\ref{lem1-forthm}, the lhs is equivalent to 
\begin{align}
L(z)&= \frac{1}{2\pi\mathrm{i}}
\int_{\Gamma_r} dw\;
\int_{\mathbb{R}} ds \frac{\varrho(s)
}{(s+w)^2}
\log \Big(1-\frac{R(-w) +\alpha w-\beta}{R(z)+\alpha w-\beta}\Big)\;.
\label{eq:L-Gamma-r}
\end{align}
As before, it is enough to prove that $L(z)$ evaluates to the rhs of
(\ref{eq:L}) for $|\lambda|<\lambda_\epsilon$ where 
the logarithm in (\ref{eq:L-Gamma-r}) can be expanded into a
uniformly convergent power series:
\begin{align*}
L(z) 
&=-\sum_{n=1}^\infty \frac{1}{2\pi\mathrm{i}n}
\int_{\Gamma_r} \!
\frac{dw}{(R(z)+\alpha w-\beta)^n}
\int_{\mathbb{R}} \!ds\;
\frac{\varrho(s)
}{(s+w)^2}
\Big({-}\lambda \int_{\mathbb{R}} \!dt\;
\frac{\varrho(t)}{t-w}\Big)^n .
\end{align*}
The $(w,s)$-integrand is integrable on
$\Gamma_r\times \mathbb{R}$ so that Fubini allows
us to change the $(w,s)$-integration order. We can also move the
$s$-integral in front of the summation over $n$.
We temporarily assume $z\notin \mathbb{R}$.
This assumption guarantees that the 
two poles at $w=-(R(z)-\beta)/\alpha$ and $w=-s$ are separated.
Both are located in the interior of $\Gamma_r$
(shaded in gray in Fig.~~\ref{fig:contour}). 
The residue theorem gives
\begin{align*}
L(z)
&=
\int_{\mathbb{R}} ds\;\varrho(s)
\frac{\partial}{\partial s}\Big[
\sum_{n=1}^\infty \frac{1}{n}
\frac{1}{(R(z)-\alpha s-\beta)^n}
\Big(-\lambda \int_{\mathbb{R}} dt\;
\frac{\varrho(t)}{t+s}\Big)^n \Big]
\\
&+
\int_{\mathbb{R}} ds\;\varrho(s)
\frac{\partial}{\partial s}
\Bigg[
\sum_{n=1}^\infty \frac{1}{n!} \Big(\frac{-\lambda}{\alpha}\Big)^n
\\
&\qquad\qquad \times \frac{\partial^{n-1}}{\partial w^{n-1}}\Big|_{w=0}
\frac{\displaystyle
\Big(\int_{\mathbb{R}} dt\;
\frac{\varrho(t)}{t
  +\frac{1}{\alpha}(R(z)-\beta)  -w}\Big)^n }{
w+\frac{1}{\alpha}(\alpha s+\beta-R(z))}
\Bigg]\;.
\end{align*}
The series are summable for $|\lambda|$ small enough
(depending on the distance between $R(z)$ and $\mathbb{R}$).
The first line of the rhs 
produces a standard logarithm, whereas the other integral
is processed with the B\"urmann formula (\ref{eq:Buermann}).
Setting $H_z(w)= \log \frac{w+\frac{1}{\alpha}(\alpha s+\beta-R(z))}{
  \frac{1}{\alpha}(\alpha s+\beta-R(z))}$
and taking the same $\phi_z(w)$ given in (\ref{phizw}),
the expression in $[ ~ ]$ equals
$H_z(g( -\frac{\lambda}{\alpha}))$, where 
$g( -\frac{\lambda}{\alpha})=-\frac{1}{\alpha} K_z$ as 
in the proof of Lemma~\ref{lem1-forthm}. We thus arrive at
\begin{align*}
L(z)
&=
\int_{\mathbb{R}} ds\;
\varrho(s)
\frac{\partial}{\partial s}\Bigg[
-\log \Bigg(1-\frac{\displaystyle
\lambda \int_{\mathbb{R}} dt\;
\frac{\varrho(t)}{t+s}}{
\alpha s+\beta-R(z)}
\Bigg)\Bigg]
\\
&+
\int_{\mathbb{R}} ds\;
\varrho(s)
\frac{\partial}{\partial s}
\Bigg[
\log \frac{\alpha s+\beta-R(z)-K_z}{
\alpha s+\beta-R(z)}
\Bigg]
\\
&= \int_{\mathbb{R}} ds\;
\varrho(s)
\frac{\partial}{\partial s}\Big[
\log \frac{s-z}{R(s)-R(z)}\Big]\;,
\end{align*}
where (\ref{def:Rnew}) and $K_z$ from the proof of
Lemma~\ref{lem1-forthm} have been used.

The final formula is the assertion; so far
for $z\notin \mathbb{R}$.  The result is continuous in $z$ in a
neighboorhood $V$ of $\mathbb{R}_{>0}$, holomorphic on
$V\cap\{\mathrm{Im}(z)>0\}$ and $V\cap\{\mathrm{Im}(z)<0\}$. By
Morera's theorem the two regions $\mathrm{Im}(z)>0$ and
$\mathrm{Im}(z)<0$ patch holomorphically together and define the same
holomorphic function on $V$.
\end{proof}
\end{lemma}

\begin{proof}[Proof of Theorem~\ref{thm:R}]
  Formula (\ref{identity-R}) is the sum of (\ref{eq:Kz}) and $\lambda$ times
  \ref{eq:L}.
\end{proof}

\begin{proposition}
For $z\in \mathcal{U} \setminus \mathbb{R}$,
$y\in \mathcal{U}\cap \mathbb{R}$ and $\lambda\in \mathcal{L}$,
consider the integral
\begin{align}
  \Psi(z;y)&:=  \frac{1}{2\pi\mathrm{i}}\int_{\Gamma}
  dw \;  \frac{R'(w) 
  }{R(w)-R(z)}\log (R(y) -R(-w))\;,
\label{def:Psi}
\end{align}
where the contour $\Gamma$ encircles the branch cut $[M^2,\Lambda^2]$
of $w\mapsto R(-w)$ but excludes $z$. This integral evaluates
for $y,\mathrm{Re}(z)>\beta$ and
$\mathrm{Im}(z)\neq 0$ to 
\begin{align}  
\Psi(z;y)
&=-\log \alpha + \log (R(y)  -R(-z))
+\log\Big(\frac{R(y)+R(z)}{(y+ R(z))(z+R(y))}\Big)
\nonumber
\\
&+  \frac{1}{2\pi\mathrm{i}}\int_{\mathbb{R}} ds 
\Big(  \frac{d}{ds} \log \Big( \frac{R(z)-R(\mathrm{i}s)}{
  R(z)-\mathrm{i}s }\Big)\Big)
  \log \Big(\frac{R(y)-R(-\mathrm{i}s)}{R(y)+\mathrm{i}s}\Big)\;.
\label{result-Psi}
\end{align}
\begin{proof}
  The function $w \mapsto \log (R(y)+w)$ is
  holomorphic in a neighbourhood of $[M^2,\Lambda^2]$ which contains
  $\Gamma$ so that by Cauchy's theorem its integral over $\Gamma$
  vanishes. We absorb it into $\Psi(z;y)$:
  \begin{align}
  \Psi(z;y)
  = \frac{1}{2\pi\mathrm{i}}\int_{\Gamma} dw\;
  \frac{R'(w)}{R(w)-R(z)} \log \Big(\frac{R(y) -R(-w)}{
    R(y)+w} \Big)\;.
\label{Psi-var}
\end{align}
We deform $\Gamma$ to a contour $\Gamma_z$ which encircles both
$[M^2,\Lambda^2]$ and the point $z\in \mathcal{U}\setminus \mathbb{R}$.
See Figure~\ref{fig:contour2}.
\begin{figure}[ht]
\begin{center}\begin{tikzpicture}
  \tkzDefPoint(-6,0){GO}
  \tkzDefPoint(-6,1){GB}  
  \tkzDefPoint(-5.5,0){GM}
  \tkzDefPoint(-5.5,0.3){GM1}
  \tkzDefPoint(-5.5,-0.3){GM2}
  \tkzDefPoint(-4.8,0){GL}
  \tkzDefPoint(-5,1){GZ}
  \tkzDefPoint(-5,1.3){GZR}  
  \tkzDefPoint(-4.95,1){GZ2}    
  \tkzDefPoint(-5.05,1){GZ1}
  \tkzDefPoint(-4.95,0){GZ4}    
  \tkzDefPoint(-5.05,0){GZ3}      
  \tkzDefPoint(-4.8,0.3){GL1}  
  \tkzDefPoint(-4.8,-0.3){GL2}
  \tkzInterLC(GZ1,GZ3)(GZ,GZR)\tkzGetPoints{GB3}{GB1}
  \tkzInterLC(GZ2,GZ4)(GZ,GZR)\tkzGetPoints{GB2}{GB4}
  \tkzInterLL(GZ1,GZ3)(GL1,GM1)\tkzGetPoint{GC1}
  \tkzInterLL(GZ2,GZ4)(GL1,GM1)\tkzGetPoint{GC2}
  \tkzDrawPoint(GZ)
  \tkzLabelPoint[above,yshift=-0.2ex](GZ){\footnotesize$z$}
  \tkzDrawLine[-Latex, add=1 and 0.5](GO,GB)
  \tkzDrawLine[-Latex, add=0.5 and 0.5](GO,GL)
  \tkzDrawLine[add=0 and 0,line width=2pt](GL,GM)
  \tkzDrawArc[line width=1pt](GL,GL2)(GL1)    
  \tkzDrawArc[line width=1pt](GM,GM1)(GM2)
  \tkzDrawLine[add=0 and 0,line width=1pt,tkz arrow={Latex[scale=0.7]}](GM1,GC1)
  \tkzDrawLine[add=0 and 0,line width=1pt](GC1,GB1)
  \tkzDrawLine[add=0 and 0,line width=1pt,tkz arrow={Latex[scale=0.7]}](GB2,GC2)
  \tkzDrawLine[add=0 and 0,line width=1pt](GC2,GL1)
  \tkzDrawLine[add=0 and 0,line width=1pt,tkz arrow={Latex[scale=0.7]}](GL2,GM2)
  \tkzDrawArc[line width=1pt](GZ,GB2)(GB1)
  \tkzDefPoint(0,0){O}
  \tkzDefPoint(0,1){B}  
  \tkzDefPoint(2.1,0){R}
  \tkzDefPoint(0.5,0){M}
  \tkzDefPoint(0.5,0.3){M1}
  \tkzDefPoint(0.5,-0.3){M2}
  \tkzDefPoint(0.5,0.07){M3}
  \tkzDefPoint(0.5,-0.07){M4}  
  \tkzDefPoint(1.2,0){L}
  \tkzDefPoint(1.2,0.3){L1}  
  \tkzDefPoint(1.2,-0.3){L2}
  \tkzDefPoint(1.2,0.07){L3}  
  \tkzDefPoint(1.2,-0.07){L4}
  \tkzInterLC(L3,M3)(O,R)\tkzGetPoints{R1}{R3}
  \tkzInterLC(L4,M4)(O,R)\tkzGetPoints{R4}{R2}  
  \tkzInterLC(L4,M4)(L,L1)\tkzGetPoints{P3}{P2}
  \tkzInterLC(L3,M3)(L,L1)\tkzGetPoints{P1}{P4}
  \tkzInterLC(M2,O)(O,R)\tkzGetPoints{S2}{S1}  
  \tkzDefPoint(1,1){Z}
  \tkzDefPoint(1,1.3){ZR}  
  \tkzDefPoint(1.05,1){Z2}    
  \tkzDefPoint(0.95,1){Z1}
  \tkzDefPoint(1.05,0){Z4}    
  \tkzDefPoint(0.95,0){Z3}      
  \tkzInterLC(Z1,Z3)(Z,ZR)\tkzGetPoints{B3}{B1}
  \tkzInterLC(Z2,Z4)(Z,ZR)\tkzGetPoints{B2}{B4}
  \tkzInterLL(Z1,Z3)(L1,M1)\tkzGetPoint{C1}
  \tkzInterLL(Z2,Z4)(L1,M1)\tkzGetPoint{C2}  %
  \tkzFillCircle[gray!20](O,R)
  \tkzFillCircle[white](L,L1)  
  \tkzFillCircle[white](M,M1)
  \tkzFillCircle[white](Z,ZR)  
  \tkzFillPolygon[white](L1,M1,M2,L2)
  \tkzFillPolygon[white](C1,B1,B2,C2)  
  \tkzFillPolygon[white](P1,R1,R2,P2)      
  \tkzDrawPoint(Z)
  \tkzLabelPoint[above,yshift=-0.2ex](Z){\footnotesize$z$}
  \tkzDrawLine[-Latex, add=2.5 and 1.6](O,B)
  \tkzDrawLine[-Latex, add=2.3 and 1.3](O,L)
  \tkzDrawLine[add=0 and 0,line width=2pt](L,M)
  \tkzDrawArc[line width=1pt](L,L2)(P2)    
  \tkzDrawArc[line width=1pt](L,P1)(L1)    
  \tkzDrawArc[line width=1pt](M,M1)(M2)
  \tkzDrawLine[add=0 and 0,line width=1pt,tkz arrow={Latex[scale=0.7]}](M1,C1)
  \tkzDrawLine[add=0 and 0,line width=1pt,tkz arrow={Latex[scale=0.7]}](L2,M2)
  \tkzDrawArc[line width=1pt,tkz arrow={Latex[scale=0.7]}](O,R1)(R2)
  \tkzDrawLine[add=0 and 0,line width=1pt,tkz arrow={Latex[scale=0.7]}](P1,R1)
  \tkzDrawLine[add=0 and 0,line width=1pt,tkz arrow={Latex[scale=0.7]}](R2,P2)
  \tkzDrawLine[add=0 and 0,line width=1pt](C1,B1)
  \tkzDrawLine[add=0 and 0,line width=1pt,tkz arrow={Latex[scale=0.7]}](B2,C2)
  \tkzDrawLine[add=0 and 0,line width=1pt](C2,L1)
  \tkzDrawArc[line width=1pt](Z,B2)(B1)  
  \tkzDrawLine[->,add=0 and 0](O,S1)
  \tkzLabelLine[above](O,S1){\small$r$}
\end{tikzpicture}\vspace*{-0.5cm}
\end{center}
\caption{Sketch of the integration contours
  $\Gamma_z$ (left) and
  $\Gamma_{z,r}^*$ (right). 
The fat part of the real axis indicates the 
interval $[M^2,\Lambda^2]$.
The contour   $\Gamma_{z,r}^{**}$ is the restriction of
$\Gamma_{z,r}^{*}$ to the half plane with non-negative real part.
\label{fig:contour2}}
\end{figure}
The difference is the residue at $z$:
\begin{align}
  &\Psi(z;y)
  \label{Psi-a}
  \\
  &= \log \Big(\frac{R(y)  -R(-z)}{R(y)+z} \Big)
  + \frac{1}{2\pi\mathrm{i}}\int_{\Gamma_z} dw\;
  \frac{R'(w)}{R(w)-R(z)} \log \Big(\frac{R(y)  -R(-w)}{R(y)
    +w} \Big)
  \nonumber
  \\
&= \log \Big(\frac{R(y) -R(-z)}{R(y)+z} \Big)
  \nonumber
  \\
  &+\frac{1}{2\pi\mathrm{i}}\int_{\Gamma_z}dw
  \Big(\frac{d}{dw}
\log \Big( \frac{R(z)-R(w)}{R(z)-w}\Big)\Big)
\log \Big(\frac{R(y)-R(-w)}{R(y)+w}\Big)
\tag{**}
\\
& + \frac{1}{2\pi\mathrm{i}}\int_{\Gamma_z}
\frac{dw}{w   -R(z)}
\Big[\log \Big(\frac{R(y)-R(-w)}{R(y)+w}\Big)
-\log\alpha +\log \alpha\Big]
\;.
\tag{*}
\end{align}
The line (*) cancels with a part of the line (**). Since we
will process these two lines differently, we assume that the
artificially introduced pole at
$w=R(z)$, which for small $|\lambda|$ is close to
$w=z$, is also contained in the interior of
$\Gamma_z$.

One has $\log \big(\frac{R(.)-R(\mp w)}{R(.)\pm w}\big)
\sim \log\alpha+\mathcal{O}(w^{-1})$ for $w\to \infty$.
In the line (**) 
of (\ref{Psi-a}) we can thus extend $\Gamma_z$ to a contour
$\Gamma_{z,r}^{**}$ which starts at $-\mathrm{i}r$ for large $r$,
goes a quarter circle to $+r$, from there along the real axis (in
negative direction) to the intersection $p$ with $\Gamma_z$,
follows $\Gamma_z$ clockwise until $p$, goes along the real
axis (now positive direction) from $p$ to $r$ and finally along a
quarter circle from $r$ to $\mathrm{i}r$
($\Gamma_{z,r}^{**}$  would be the restriction of the right part of
Figure~\ref{fig:contour2} to non-negative real part).
The additional parts from $r$
to $p$ and $p$ to $r$ cancel, and for $r\to \infty$ the integral over
the quarter circles vanishes because of
$\frac{d}{dw}
\log \big( \frac{R(z)-R(w)}{R(z)-w}\big)
=\mathcal{O}(w^{-2})$ for $w\to \infty$.
We can then deform the contour 
$\Gamma_{z,r}^{**}$ to the straight line $\mathrm{i}\mathbb{R}$.
Since no poles or branch cuts are crossed by the deformation, the integral in the
line (**) is unchanged when replacing $\Gamma_z$ by
$\mathrm{i}\mathbb{R}$. Setting $\mathrm{i}\mathbb{R}\ni w=\mathrm{i}s$,
we thus recover the last line of (\ref{result-Psi}).

In the line (*) of  (\ref{Psi-a}), the integral of the
final term $+\log\alpha$ inside $[\dots]$ is $-\log\alpha$
(note that $\Gamma_z$ encircles the pole at $w=R(z)$ in negative orientation).
In the remainder which we denote by
$\Psi_{*}(z;y)$ we are allowed to 
extend the contour $\Gamma_z$ to
$\Gamma_{z,r}^*$ obtained by connecting the end point
$\pm \mathrm{i}r$ of $\Gamma_{z,r}^{**}$ by a half circle of radius $r$
in the plane $\mathrm{Re}(w)<0$ (sketched in the
right part of Figure~\ref{fig:contour2}).
The integral over the circle
vanishes for $r\to \infty$. We will prove
\begin{align}
  \Psi_{*}(z,y)&:=
\frac{1}{2\pi\mathrm{i}}\int_{\Gamma_{z,r}^*}
\frac{dw}{w   -R(z)}
\log \Big(\frac{R(y)-R(-w)}{\alpha(R(y)+w)}\Big)
\nonumber
\\
&=
  \log \Big(\frac{R(y)+R(z)}{R(z)+y}
\Big)  \;,
\label{Psi-b}
\end{align}
and this (and the previous discussion) brings (\ref{Psi-a}) 
into the assertion (\ref{result-Psi}). Both sides of
(\ref{Psi-b}) are holomorphic in $\lambda \in \mathcal{L}$.
It is thus enough to prove (\ref{Psi-b})
for $|\lambda|<\lambda_\epsilon$ where
$\lambda_\epsilon$ is such that the logarithm in $\Psi_*(z,y)$
can be expanded into a uniformly convergent power series:
\begin{align*}
  \Psi_{*}(z;y)
&=-\sum_{n=1}^\infty \frac{(-\lambda/\alpha)^n}{2\pi\mathrm{i}n}
\int_{\Gamma_{z,r}^*}
\frac{dw}{(w   -R(z))(R(y)+w)^n }
\nonumber
\\*
&\qquad\qquad \times \Big(
\frac{(1-\alpha)}{\lambda}R(y)-\frac{\beta}{\lambda}
+
\int_{\mathbb{R}} dt \frac{\varrho(t)}{t-w}
\Big)^n.
\end{align*}
In the interior of the region bordered by $\Gamma_{z,r}^*$ (shaded in gray in
Figure~\ref{fig:contour2}),
the integrand has a pole of order $n$ at $w=-R(y)$,
whereas the poles at
$w   =R(z)$ and $w=t$ are outside of
$\Gamma_{z,r}^*$. The residue theorem gives 
\begin{align*}
  \Psi_{*}(z;y)
&=\sum_{n=1}^\infty \frac{(-\lambda/\alpha)^n}{n!}
\frac{d^{n-1}}{dw^{n-1}}\Big|_{w=0}
\Big[
\frac{d}{dw}\log \Big(\frac{R(y)+R(z)}{
   R(y)+R(z)-w}
 \Big)
\\
&\qquad\qquad\times
\Big(
\frac{1-\alpha}{\lambda}R(y)-\frac{\beta}{\lambda}
+\int dt \frac{\varrho(t)}{t+R(y)-w}
\Big)^n\Big]\;.
\end{align*}
The integral is of the form of the B\"urmann formula
(\ref{eq:Buermann})
for $z\mapsto -\lambda/\alpha$ and 
$\phi_y(w)=\frac{1-\alpha}{\lambda}R(y)-\frac{\beta}{\lambda}
+\int_{\mathbb{R}} dt\;
\frac{\varrho(t)}{
  t +R(y)  -w}$ as well as 
$H_{y,z}(w)= \log \big(\frac{R(y)+R(z)}{R(y)+R(z)-w}\big)$.
We thus consider the auxiliary integral 
\begin{align*}
  \Psi^{(0)}(y)
&=\sum_{n=1}^\infty \frac{(-\lambda/\alpha)^n}{n!}
\frac{d^{n-1}}{dw^{n-1}}\Big|_{w=0}
\Big(\frac{1-\alpha}{\lambda}R(y)-\frac{\beta}{\lambda}
+\int dt \frac{\varrho(t)}{t+R(y)-w}
 \Big)^n
\end{align*}
for which the Lagrange inversion formula gives
\[
  -\frac{\lambda}{\alpha}
  =  \frac{\Psi^{(0)}(y)}{\phi_y(\Psi^{(0)}(y))}
\]
or
\[
  (\alpha-1)R(y)+\beta
  - \lambda
\int dt \frac{\varrho(t)}{t+R(y)-\Psi^{(0)}_*(y)}
=\alpha \Psi^{(0)}(y)\;.
\]
This amounts to $R(R(y)-  \Psi^{(0)}(y))=R(y)$.
For small enough $|\lambda|$,
$R(y)- \Psi^{(0)}(y)$ is close to $R(y)$, in particular in $H_+$ where $R$ is injective. This means
$\Psi^{(0)}(y)=R(y)-y$. With this auxiliary result the B\"urmann formula
(\ref{eq:Buermann}) gives
\[
\Psi_{*}(z;y)= 
\log \Big(\frac{R(y)+R(z)}{R(y)+R(z)-\Psi^{(0)}(y)}\Big)
=\log \Big(\frac{R(y)+R(z)}{y+R(z)}\Big)\;.
\]
We thus confirm
(\ref{Psi-b}), 
first for small $|\lambda|$,
but then for all $\lambda\in \mathcal{L}$ by holomorphicity.
Everything together proves the assertion
(\ref{result-Psi}).
\end{proof}
\end{proposition}

\subsection{The 2-point function}

\label{sec:G}

We follow a strategy explained e.g.\ in sections 4.2 and 4.4 of
Tricomi's classical book \cite{Tricomi:1957??} where a 
theorem due to Titchmarch is the key step:

\begin{theorem}[\cite{Titchmarsh:1937kpd}, Thm 103]
  \label{thm:Titchmarch}
  Let $\Phi:\mathbb{H}\to \mathbb{C}$ be analytic on the upper
  half plane $\mathbb{H}=\{z\in \mathbb{C}\;|~~
  \mathrm{Im}(z)>0\}$ such that
  \begin{align}
\int_{\mathbb{R}} dx\;|\Phi(x+\mathrm{i}y)|^p
  \leq  K \qquad \text{for any  $y>0$,   for some $p>1$ and some $K$}.
\label{bound-Titchmarch}
\end{align}
Then $\lim_{\epsilon\searrow 0} \Phi(x+\mathrm{i}\epsilon)=:u(x)+\mathrm{i}v(x)$
exists, and the real-valued functions
$u,v\in L^p(\mathbb{R})$ are almost everywhere related by 
\[
 \frac{1}{\pi} \fint_{\mathbb{R}} \frac{u(t)dt}{t-x} =v(x)\;,\qquad 
 \frac{1}{\pi} \fint_{\mathbb{R}} \frac{v(t)dt}{t-x} =u(x)\;.
\]
\end{theorem}

For the following steps we assume that the measure $\varrho$
in (\ref{def:Rnew}) safisfies the 
  Sokhotski-Plemelj theorem
\cite{Sokhotski:1873, Plemelj:1908},
    \begin{align}
    \lim_{\epsilon\searrow 0} \mathrm{Im}
    \Big(
    \frac{1}{\pi}
    \int_{\mathbb{R}}dt \;\frac{\varrho(t)}{t-(x+\mathrm{i}\epsilon)}
\Big)&=   \varrho(x)\;,
\label{Plemelj}
\\
    \lim_{\epsilon\searrow 0} \mathrm{Re}
\Big(\int_{\mathbb{R}}dt \;\frac{\varrho(t)}{t-(x+\mathrm{i}\epsilon)}
\Big)
&=   \fint_{\mathbb{R}} dt\;
\frac{\varrho(t)}{t-x}\;,
\nonumber
\end{align}
where $\fint$ is the Cauchy principal value integral. It is well-known
(see e.g.\ \cite{Muskhelishvili})
that the Plemelj formulae (\ref{Plemelj}) hold for H\"older-continuous
functions $\varrho$. Slightly weaker regularity suffices
  \cite{castillo2024pointwiseexistencecauchyrmpv}.
  For such $\varrho$ and corresponding function $R$, we define for $x,y>0$
  the boundary value
\begin{align}
  \tau(x;y):=\lim_{\epsilon\searrow 0}
  \mathrm{Im}\big(\log (R(y)-R(-x-\mathrm{i}\epsilon))\big)\;.
\label{tau-R}
\end{align}
Taking the Plemelj formula $\lim_{\epsilon\searrow 0}
\mathrm{Im}(-R(-x-\mathrm{i}\epsilon)) =\lambda\pi \varrho(x)$ into account,
we get
\[
  R(y)-\lim_{\epsilon\searrow 0} \mathrm{Re} (R(-x-\mathrm{i}\epsilon))
  = \lambda\pi \varrho(x) \cot \tau(x;y)
\]
and then
\begin{align}
\lim_{\epsilon\searrow 0}  \log (R(y)-R(-x-\mathrm{i}\epsilon))
= \mathrm{i}\tau(x;y)+ \log \Big(\frac{\lambda\pi \varrho(x)}{
  \sin \tau(x;y)}\Big)\;.
\end{align}

Consider now for $z\in R(\mathcal{U}\setminus \mathbb{R})$ and
$y\in R(\mathcal{U}\cap \mathbb{R})$ the function
\begin{align}
  \Phi(z;y):=\exp(\Psi(R^{-1}(z);R^{-1}(y)))-1\;,
\label{def:Phi}
\end{align}
where $\Psi$ was introduced in (\ref{def:Psi}).
From (\ref{result-Psi}) and the above discussion we get the limit
\begin{align}
  \lim_{\epsilon \searrow 0}  \Phi(x+\mathrm{i}\epsilon;y)
  &= \mathrm{i}\lambda \pi \varrho(R^{-1}(x)) \alpha^{-1}G(x,y)
  \nonumber
  \\
  & +
  \lambda \pi \varrho(R^{-1}(x)) \alpha^{-1} G(x,y)
  \cot \tau(R^{-1}(x);R^{-1}(y))-1\;,
\label{limPhi}
\end{align}
if $R(x)\in \mathrm{supp}(\varrho)$, where
\begin{align}
  G(x,y)&:=
  \frac{(x+y) \exp\Big[\displaystyle
\frac{1}{2\pi\mathrm{i}}\int_{\mathbb{R}} ds 
\Big(  \frac{d}{ds} \log \Big( \frac{x-R(\mathrm{i}s)}{
  x-\mathrm{i}s}\Big)\Big)
  \log \Big(\frac{y-R(-\mathrm{i}s)}{y+\mathrm{i}s}\Big)\Big]}{
  (y+ R^{-1}(x))(x+ R^{-1}(y))}\;.
  \label{G-formula}
\end{align}
Note that, via integration by parts, the integral inside $[~]$
is real and symmetric in $x,y$, and so is $G(x,y)$.

For $z=R^{-1}(x+\mathrm{i}s)$ away from $\Gamma$, the definition
(\ref{def:Psi}) of $\Psi$ shows that
$x+\mathrm{i}s\mapsto \Phi(x+\mathrm{i}s,y)$ is $L^p$ and satisfies
the bound (\ref{bound-Titchmarch}) of Thm.~\ref{thm:Titchmarch}
for these $x+\mathrm{i}s$. We stress
that the $L^p$-condition fixes the final term $-1$ in
(\ref{def:Phi}). When $x+\mathrm{i}s$ approaches
$[R(M^2),R(\Lambda^2)]$, we need some $L^p$-existence of the limit
(\ref{limPhi}) to guarantee that $\Phi$ remains globally $L^p$ on the
upper half plane with a uniform bound (\ref{bound-Titchmarch}). For
that it is enough
that $\varrho$ is H\"older-continuous. Under such
assumptions, Thm.\ \ref{thm:Titchmarch} states that the Hilbert
transform of the imaginary part of
$\lim_{\epsilon \searrow 0} \Phi(x+\mathrm{i}\epsilon;y)$ equals
(almost everywhere) its real part:
\begin{align*}
&1+ \lambda \fint_{\mathbb{R}} dt\;
\frac{\varrho(R^{-1}(t)) \alpha^{-1} G(t,y)}{t-x}
\\
&=\alpha^{-1} G(x,y) \lambda\pi \varrho_{,\Lambda}(R^{-1}(x))
\cot \tau(R^{-1}(x);R^{-1}(y))
\nonumber
\\
&\equiv \alpha^{-1} G(x,y) \Big(y-\lim_{\epsilon\searrow 0}
\mathrm{Re}(R(-R^{-1}(x+\mathrm{i}\epsilon))\Big)\;.
\end{align*}

In the final line we insert our result (\ref{identity-R})
at $z=R^{-1}(x+\mathrm{i}\epsilon)$:
\begin{align}
&1+ \lambda \fint_{\mathbb{R}} dt\;
\frac{\varrho(R^{-1}(t)) \alpha^{-1} G(t,y)}{t-x}
\\
&= \alpha^{-1} G(x,y) \Big(y+ x -2\beta
+  \frac{1}{2\pi\mathrm{i}}
\int_{\Gamma} dw\;R'(w)
\log \big(x- R(-w)\big)
\nonumber
\\
&\qquad\qquad +\lim_{\epsilon\searrow 0}
\mathrm{Re}\Big(
\lambda\int_{\mathbb{R}} dt\;
\frac{\varrho(R^{-1}(t))}{t-(x+\mathrm{i}\epsilon)}
\Big)\Big)\;.
\nonumber
\end{align}
The real part of the final integral is the principal value. We move it to
the lhs and notice that since $G$ is real-analytic in the first argument,
the principal value integral of the resulting difference quotient
converges to the ordinary integral. The contour integral over $\Gamma$ is
reexpressed via (\ref{def:Psi})  for $y\mapsto R^{-1}(x)$
and $z\mapsto R^{-1}(s)$ with $s>R(\Lambda^2)$ large:
\begin{align*}
 \frac{1}{2\pi\mathrm{i}}
\int_{\Gamma} dw\;R'(w)
\log \big(x- R(-w)\big)
 & = -\lim_{s\to \infty}
s \Psi(R^{-1}(s),R^{-1}(x))
\\
&= -\lim_{s\to \infty} s \Phi(s,x)\;.
\end{align*}
But $\Phi(s,x)$ is real for large $s$, and this real part is the
Hilbert transform of the imaginary part:
\begin{align}
 \frac{1}{2\pi\mathrm{i}}
\int_{\Gamma} dw\;R'(w)
\log \big(x- R(-w)\big)
&=- \lim_{s\to \infty} \frac{s}{\pi}
\fint_{\mathbb{R}} dt \frac{\lim_{\epsilon\searrow 0} \mathrm{Im}(\Phi(
  t+\mathrm{i}\epsilon))}{t-s}
\nonumber
\\
&=\frac{1}{\pi}\int_{\mathbb{R}} dt \;\lim_{\epsilon\searrow 0} \mathrm{Im}(\Phi(
t+\mathrm{i}\epsilon))
\nonumber
\\
&=\lambda\int_{\mathbb{R}} dt \;\varrho(R^{-1}(t)) \alpha^{-1} G(t,x)\;.
\label{Psi-G}
\end{align}
We have thus proved:
\begin{theorem}
\label{thm:final}
Starting from the function $R$ defined in (\ref{def:Rnew}),
with $\varrho$ satisfying the Sokhotski-Plemelj theorem (\ref{Plemelj}),
the part
\begin{align*}
  G(x,y)&:=
  \frac{(x+y) \exp\Big[\displaystyle
\frac{1}{2\pi\mathrm{i}}\int_{\mathbb{R}} ds 
\Big(  \frac{d}{ds} \log \Big( \frac{x-R(\mathrm{i}s)}{
  x-\mathrm{i}s}\Big)\Big)
  \log \Big(\frac{y-R(-\mathrm{i}s)}{y+\mathrm{i}s}\Big)\Big]}{
  (y+ R^{-1}(x))(x+ R^{-1}(y))}
\end{align*}
given in (\ref{G-formula}) of the boundary value (\ref{limPhi})
fulfils the non-linear integral equation
\begin{align}
&1+ \lambda \int_{\mathbb{R}} dt\;
\varrho(R^{-1}(t)) \frac{\alpha^{-1} G(t,y)-\alpha^{-1} G(x,y)}{t-x}
\nonumber
\\
&= \alpha^{-1} G(x,y) \Big(y+ x -2\beta
+ 
\lambda\int_{\mathbb{R}} dt \;\varrho(R^{-1}(t)) \alpha^{-1} G(t,x)
\Big)\;.
\end{align}      
\end{theorem}

Comparing with (\ref{eq:Gint}) we have established
a solution of the initial equation for the 2-point function (a loop equation or
Dyson-Schwinger equation) of a quartic matrix
model if we identify
\begin{align}
  \rho_0(x)=\varrho(R^{-1}(x))\;,\qquad
Z=\alpha^{-1}\;,\qquad \mu_{bare}^2=-2\beta\;.
\end{align}
The meaning of the parameters $Z=\alpha^{-1}$ and $\mu_{bare}^2=-2\beta$ will be
discussed in sec.~\ref{sec:renormalisation}.  Since $R$ also
contains $\varrho$, it is still some
challenge to solve for given measure $\rho_0$ and given parameters
$\alpha,\beta$ the resulting integral equation
\begin{align}
  \rho_0\Big(\alpha x+\beta-\lambda
  \int_{\mathbb{R}} dt\;\frac{\varrho(t)}{t+x}\Big)
=\varrho(x)\;.
\label{relation-rho}
\end{align}
We will discuss in sec.~\ref{sec:ex} three important cases for $\rho_0$
where a solution has been achieved. The converse interpretation that one
chooses $\varrho$ and
defines a corresponding quartic matrix model by the measure
$\rho_0(x)=\varrho(R^{-1}(x))$ is easy: Every
Herglotz-Nevanlinna function $y(z)=-R(-z)$ for which the measure has support
in $[M^2,\infty)$ defines a unique quartic matrix model.

\subsection{Renormalisation}
\label{sec:renormalisation}

As long as a single $\varrho$ is considered we can make
any choice of $\alpha, \beta$; the simplest one being $\alpha=1$ and
$\beta=0$. The parameters are relevant if we consider families
$\varrho$ where the upper bound $\Lambda^2$ of the support
of the measure goes to $\infty$. Then the  integral (\ref{def:Rnew})  
might diverge. Depending on the rate at which
$\varrho(t)$ grows with $t$, special functional
dependencies of $\alpha,\beta$ on $\Lambda$ will be necessary
to define $R(z)$ in the limit $\Lambda\to \infty$.
\begin{definition}
\label{def:dimension}
The spectral dimension\footnote{This definition captures Weyl's law
\cite{Weyl:1911} of the asymptotics of eigenvalues of the Laplacian.}
of a spectral measure function $f$
(e.g.\ $f=\rho_0$ or $f=\varrho$) 
is defined by $D_{spec}(f) := 
\inf\{p\;|~ \int_0^\infty
\frac{dt\;f(t)}{(t+1)^{p/2}}\text{ converges}\}$. The
renormalisation procedure is classified by the number 
$D=2[\frac{1}{2} D_{spec}(f)] \in \{0,2,4,> 4\}$ as follows:
\begin{description}
\item[$D=0$:] One can set $Z=\alpha^{-1}$ and $\mu_{bare}^2=-2\beta$ to any
  finite value, e.g.\ $Z=1$, $\mu^2_{bare}=0$.

\item[$D=2$:] 
One can set $Z=\alpha^{-1}$ to any finite value (e.g.\ $Z=1$), but 
$\mu^2_{bare}(\Lambda)=-2\beta$ diverges with $\Lambda^2$. 
The simplest choice\footnote{One could also take
  $\beta=\beta_0+\lambda \int_{\mathbb{R}} dt\;\frac{\varrho(t)}{
    t+\beta_1}$ for some $\beta_0,\beta_1$. The asymptotic behavior of
  $\beta(\Lambda)$ is fixed; in the subleading contributions there is a
  certain freedom.} is Taylor subtraction
\begin{align}
  \beta=\Big(\lambda \int_{\mathbb{R}} dt\;\frac{\varrho(t)}{t+\mu}
  \Big)_{\mu=0}
\quad\Rightarrow\quad    
R(z)=z +
\lambda z \int_{\mathbb{R}} dt\;\frac{\varrho(t)}{t(t+z)}\;.
\label{Taylor-2}
\end{align}

\item[$D=4$:] Both
  $\mu^2_{bare}(\Lambda)=-2\beta$ and $Z(\Lambda)=\alpha^{-1}$ diverge
  for $\Lambda\to \infty$. The simplest choice is Taylor subtraction
\begin{align}
  \beta&=\Big(\lambda \int_{\mathbb{R}} dt\;\frac{\varrho(t)}{t+\mu}
  \Big)_{\mu=0}\;, \qquad
  \alpha=1+\Big(\frac{\partial}{\partial \mu}
  \lambda \int_{\mathbb{R}} dt\;\frac{\varrho(t)}{t+\mu}
  \Big)_{\mu=0}
  \label{Taylor-4}
  \\
  &\Rightarrow\qquad    
R(z)=z -
\lambda z^2 \int_{\mathbb{R}} dt\;\frac{\varrho(t)}{t^2(t+z)}\;.
\nonumber
\end{align}

\item[$D>4$:] 
This case cannot be renormalised anymore.
\end{description}

\end{definition}
Note that the support of $\varrho$ starts at $M^2>0$ so
that there is no divergence at $t=0$ in (\ref{Taylor-2}) and
(\ref{Taylor-4}). Both equations for $R$ and the case $D=0$ can be
combined to
\begin{align}
  R(z)=z-\lambda (-z)^{D/2}
\int_{\mathbb{R}} dt\;\frac{\varrho(t)}{t^{D/2}(t+z)}\;.
\end{align}

\begin{remark}
  Recall the standard representation
  \[
az+b+\int _{\mathbb {R} } \Big(\frac {1}{t -z}-\frac {t}{1+t^2}\Big)
  d \varrho(t) \;,\qquad z\in \mathbb{H}\;,
\]
(with $b$ real, $a\geq 0$ and $\varrho$ a Borel measure on $\mathbb{R}$
satisfying
$\int _{\mathbb {R} } \frac{d \varrho(t)}{1+t^{2}}<\infty$)
of a Herglotz-Nevanlinna function. The correction term
$-\frac {t}{1+t^2}$ has the  same purpose as our choice of $\beta$ for $D=2$,
to achieve convergence of  the integral for (half-) infinite support
of the measure. It is an early example of renormalisation.
\hfill $\triangleleft$
\end{remark}

\begin{remark}
  In the proof we decisively used the injectivity of $R$; also the
  result (\ref{G-formula}) for the 2-point function
  involves $R^{-1}$ und thus relies on injectivity. Consider
  $\varrho(t)=t \chi_{[M^2,\Lambda^2]}$ which has
    spectral dimension $D=4$. Then with the choice
    (\ref{Taylor-4}) we have    
\begin{align*}
  R'(z)&=1-\lambda \int_{M^2}^{\Lambda^2} \frac{dt}{t}
  +\lambda \int_{M^2}^{\Lambda^2} \frac{tdt}{(t+z)^2}
  \\
  &=1-\lambda \log\Big(\frac{\Lambda^2(M^2+z)}{M^2(\Lambda^2+z)}\Big)
  + \frac{\lambda z}{\Lambda^2+z}
  - \frac{\lambda z}{M^2+z}\;.
\end{align*}
For $z=\Lambda^2\gg M^2$ we see that injectivity can only hold up to a
scale $\Lambda^2\approx \frac{2}{\sqrt{e}}M^2 e^{\frac{1}{\lambda}}$.
This is a manifestation of the \emph{Landau pole}, a severe threat for
quantum field theories in four dimensions. Conversely, in order to
admit arbitrary large scales $\Lambda$, the coupling constant
$\lambda$ must be zero. This is the infamous \emph{triviality problem}
of 4D QFT \cite{Aizenman:2019yuo}.

It seems at first sight that the quartic matrix model runs in
dimension $D=4$ into the same triviality problem. We showed in
\cite{Grosse:2019qps} that for the most interesting choice of the
measure $\rho_0(t)=t \chi_{[\tilde{M}^2,\tilde{\Lambda}^2]}(t)$ in the
    (Dyson-Schwinger) equation (\ref{eq:Gint}) for the planar 2-point
    function, \emph{the triviality problem does not occur}. The reason is
  that we can solve in this case the relation (\ref{relation-rho})
  exactly, and the resulting measure $\varrho$ for the
  auxiliary function $R$ lives effectively in spectral dimension
  $4-\frac{2}{\pi}\arcsin(\lambda\pi)$. We give some details in
  sec.~\ref{sec:Moyal}.
\hfill $\triangleleft$\end{remark}

\section{Finite matrices}

\label{sec:finiteN}

\subsection{The auxiliary functions $R$ and $\Psi$}

In this section we consider the case of finite matrices defined by
a Dirac measure (\ref{rhot}),
\begin{align}
\rho_0(t)=\frac{1}{{N}}\sum_{k=1}^d r_k \delta(t-e_k)\;.
\end{align}
Here $0<e_1<e_2<\dots<e_d$ are the pairwise different eigenvalues of $E$ and 
$r_1,\dots,r_d$ are their multiplicities, with 
$\sum_{k=1}^d r_k={N}$. We have implemented $\mu_{bare}^2=-2\beta=0$.
The consistency relation (\ref{relation-rho}) then reads
\begin{align}
  \varrho (x)
&=\frac{1}{{N}}\sum_{k=1}^d r_k \delta(R(x)-e_k)
=\frac{1}{{N}} \sum_{k=1}^d \frac{r_k}{R'(R^{-1}(e_k))} \delta(x-
R^{-1}(e_k))\;.
\end{align}
Of course, this $\varrho$ is not a H\"older-continuous function, We have to
use some approximate $\delta$-functions such as
$\delta_{\kappa}(x-x_0)=\frac{1}{\sqrt{2\pi \kappa}}
\exp(-\frac{1}{2\kappa} (x-x_0)^2)$, which is H\"older.
The resulting $\varrho\mapsto \varrho_\kappa$ should then be multiplied
by the characteristic function of $[M^2,\Lambda^2]$,
Thm.~\ref{thm:final} then holds for any $\kappa>0$, and the
usual dominated convergence proof of approximate Dirac functions
establishes the solution in the limit 
$\kappa\to 0$ to true $\delta$-distributions.

With these considerations,  $R$ takes with
(\ref{def:Rnew}) and for $\alpha=1$ and $\beta=0$
the form
\begin{align}
R(z)=z-\frac{\lambda}{{N}} \sum_{k=1}^d 
\frac{\varrho_k}{\varepsilon_k+z}\;,\qquad 
\varrho_k:= \frac{r_k}{R'(R^{-1}(e_k))}\;,\quad
\varepsilon_k := R^{-1}(e_k)\;.
\label{Rx-N}
\end{align}
This equation and its derivative evaluated at $z_l=R^{-1}(e_l)=\varepsilon_l$ 
for $l=1,\dots,d$ provide a system of $2d$ equations for the $2d$ parameters 
$\{\varepsilon_k,\varrho_k\}$:
\begin{align}
e_l  &= \varepsilon_l 
-\frac{\lambda}{{N}} \sum_{k=1}^d \frac{\varrho_k}{
\varepsilon_k +\varepsilon_l}\;, &
1 &= \frac{r_l}{\varrho_l} 
-\frac{\lambda}{{N}} \sum_{k=1}^d 
\frac{\varrho_k}{(\varepsilon_k+\varepsilon_l)^2}\;.
\label{rho-finite}
\end{align}
The implicit function theorem guarantees a solution in an 
open $\lambda$-interval, and one explicitly constructs a sequence 
converging to the solution $\{\varepsilon_k,\varrho_k\}$. 
Alternatively, (\ref{rho-finite}) can be interpreted as a system of 
$2d$ polynomial equations  ($d$ of them 
of degree $d+1$, the other $d$ of degree $2d+1$). Such systems 
have many solutions, and they will indeed be needed in intermediate
steps. The correct solution
is the one which for $\lambda\to 0$ converges to $\{e_k,r_k\}$. 

From (\ref{Rx-N}) we deduce a representation
\begin{align}
R(w)-R(z)
=(w-z) \prod_{k=1}^d \frac{w-\hat{z}^k}{w+\varepsilon_k}\;.
\label{R-rational}
\end{align}  
Here $\hat{z}^1,\dots,\hat{z}^d$ are the other preimages of
$R(z)$ under $R$; they are functions of $z$ and the initial data
$E,\lambda$. For real $z$ it follows from the intermediate value theorem 
that these preimages are interlaced between the poles
$\{-\varepsilon_k\}$ of $R$. In particular, for $z\geq 0$ and
$\lambda>0$ all $\hat{z}^k$ are real and located in 
$-\varepsilon_{k+1} < \hat{z}^k < -\varepsilon_{k}$ for $k=1,\dots d-1$
and $\hat{z}^d < -\varepsilon_d$.

In the case of isolated poles we can evaluate the integral
$\Psi(z;y)$ directly:
\begin{proposition}
\label{prop:Psi}
  For $R$ given by (\ref{Rx-N}), the integral (\ref{def:Psi})
  evaluates for $z\in \mathcal{U}\setminus \mathbb{R}$ to
\begin{align}
  \Psi(z;y)
&=\log \Big(\frac{R(-z)-R(y)}{R(-y)-R(z)}\Big)
+\sum_{k=1}^d\log \Big(\frac{R(-\hat{z}^k)-R(y)}{R(\varepsilon_k)-R(y)}
\Big)\;.\label{Phi-finite}
\end{align}
The function $y\mapsto \exp(\Psi(z;y))$ is holomorphic in a
neighbourhood of $\mathbb{R}_+$.
\begin{proof}
We insert the identity 
\begin{align}
  \frac{R'(w)}{R(w)-R(z)}
  =\frac{\partial}{\partial w} \log(R(w)-R(z))
  =\frac{1}{w-z}+ \sum_{k=1}^d \frac{1}{w-\hat{z}^k}
- \sum_{k=1}^d \frac{1}{w+\varepsilon_k}
\label{R-rational-1}
\end{align}
resulting from (\ref{R-rational})
into (\ref{Psi-var}) and take $\alpha=1$ and $\beta=0$ into account:
\begin{align*}
  \Psi(z;y)&=  \frac{1}{2\pi\mathrm{i}}\int_{\Gamma_r}
  dw \;  \Big(
\frac{1}{w-z}+ \sum_{k=1}^d \frac{1}{w-\hat{z}^k}
- \sum_{k=1}^d \frac{1}{w+\varepsilon_k}\Big)
\nonumber
\\
&\qquad\qquad\qquad \times
\log \Big(1+ \frac{\frac{\lambda}{N}
  \sum_{k=1}^d \frac{\varrho_k}{\varepsilon_k-w}}{w+R(y)}\Big)\;.
\end{align*}
We recall that the primary contour $\Gamma$ in (\ref{def:Phi})
separates the real interval
$[\varepsilon_1,\varepsilon_d]$ that contains the support
of $\varrho$ from $z$. As in the proof of Lemma \ref{lem1-forthm} we have extended $\Gamma$ to a contour $\Gamma_r$ sketched in the right part of
Figure~\ref{fig:contour}. In any case, $w\in \Gamma_r$ passes the
$\varepsilon_k$ at a certain distance. There is then a $\lambda_\epsilon>0$
such that the logarithm expands for $|\lambda|<\lambda_\epsilon$
into a power series which converges
uniformly on $\Gamma_r$: 
\begin{align*}
  \Psi(z;y)&= -\sum_{n=1}^\infty \frac{1}{2\pi\mathrm{i}n}
  \int_{\Gamma_r}
  dw \;  \Big(
\frac{1}{w-z}+ \sum_{k=1}^d \frac{1}{w-\hat{z}^k}
- \sum_{k=1}^d \frac{1}{w+\varepsilon_k}\Big)
\nonumber
\\
&\qquad\qquad\qquad \times
\Big(-\frac{\frac{\lambda}{N} \sum_{k=1}^d \frac{\varrho_k}{\varepsilon_k-w}}{w+R(y)}\Big)^n\;.
\end{align*}
We evaluate this integral by the residue theorem.
In the interior of $\Gamma_r$ we have the 
simple poles at $w=z$ and $w=\hat{z}^k$ (from the 
previous $R(w)=R(z)$), the simple poles at $w=-\varepsilon_k$
(from the previous $R'(w)$)
and the $n$-fold pole at $w=-R(y)$.
The $n$-fold pole at $w=\varepsilon_k$ is located outside $\Gamma_r$ and
does not contribute:
\begin{align}
  &  \Psi(z;y)
  \\
  &= -\sum_{n=1}^\infty \frac{1}{n}
\Big\{
\Big(-\frac{\frac{\lambda}{N} \sum_{l=1}^d \frac{\varrho_l}{
    \varepsilon_l-z}}{z+R(y)}\Big)^n
+\sum_{k=1}^d
\Big(-\frac{\frac{\lambda}{N} \sum_{l=1}^d
  \frac{\varrho_l}{\varepsilon_l-\hat{z}^k}}{  \hat{z}^k+R(y)}\Big)^n
\nonumber
\\
&-\sum_{k=1}^d
\Big(-\frac{\frac{\lambda}{N} \sum_{l=1}^d \frac{\varrho_l}{
    \varepsilon_l+\varepsilon_k}}{  R(y)-\varepsilon_k}\Big)^n\Big\}
\nonumber
\\
&
-\frac{(-\frac{\lambda}{N})^n}{n!}
\frac{\partial^{n-1}}{\partial w^{n-1}}
\Big|_{w=-R(y)}
\Big[  \Big(
  \frac{1}{w{-}z}+ \sum_{k=1}^d \frac{1}{w{-}\hat{z}^k}
  - \sum_{k=1}^d \frac{1}{w{+}\varepsilon_k}\Big)
  \Big(\sum_{k=1}^d \frac{\varrho_k}{\varepsilon_k{-}w}
\Big)^n
\Big]
\nonumber
\\
&=\log \Big(\frac{R(y)-R(-z)}{z+R(y)}\Big)
+\sum_{k=1}^d\log \Big(\frac{R(y)-R(-\hat{z}^k)}{\hat{z}^k+R(y)}\Big)
-\sum_{k=1}^d\log \Big(\frac{R(y)-R(\varepsilon_k)}{R(y)-\varepsilon_k}\Big)
\nonumber
\\
&-\sum_{n=1}^\infty \frac{(-\lambda/N)^n}{n!}
\frac{\partial^{n-1}}{\partial w^{n-1}}
\Big|_{w=0}
\Big[  \frac{\partial H_{z;y}(w)}{\partial w}
\big(\phi_y(w)\big)^n
\Big]\;,\qquad\text{where}
\tag{*}
\\
&
H_{z;y}(w):=\log\Big(\frac{R(w-R(y))-R(z)}{R(-R(y))-R(z)}\Big)
\quad\text{and}\quad
\phi_y(w):=\sum_{k=1}^d \frac{\varrho_k}{\varepsilon_k  +R(y)-w}
\;.
\nonumber
\end{align}
We have resummed the first three series to logarithms and
reverted the decomposition (\ref{R-rational-1}) for the last series.
According to the B\"urmann formula (\ref{eq:Buermann}), the line
(*) equals $-H_{z;y}(M_y(-\lambda/N))$, where $M_y$ solves
\[
  -\frac{\lambda}{N}=\frac{M_y(-\lambda/N)}{\phi_{z;y}( M_y(-\lambda/N))}
  \quad\Leftrightarrow\quad
  R(R(y)-M_y(-\lambda/N))=R(y)\;.
\]
As before, for $|\lambda|$ small enough,
$R(y)-M_y(-\lambda/N)$ is close to
$R(y)$ and thus contained in $H_+$ where
$R$ is injective. This means $M_y(-\lambda/N)=R(y)-y$.
Putting everything together and taking (\ref{R-rational})
for $w\mapsto -R(y)$ into account, we arrive at
the assertion (\ref{Phi-finite}) -- first for small
 $|\lambda|$, then by holomorphicity for $\lambda\in \mathcal{L}$. 

Note that $R(-y)$ in  (\ref{Phi-finite}) has a pole at
$y=\varepsilon_k$, which is canceled by the zero of
$R(\varepsilon_k)-R(y)$, making
$(R(-y)-R(z))(R(\varepsilon_k)-R(y))$ holomorphic at $y=\varepsilon_k$.
Other potential poles of $y\mapsto \exp(\Psi(z,y))$ at the preimages
$\widehat{\varepsilon_k}^l$ have negative real part, and poles
related to $z\notin\mathbb{R}_+$ have non-vanishing imaginary part.  
\end{proof}
\end{proposition}

\subsection{The 2-point function}

 The ramified covering $R$ is biholomorphic in a neighbourhood of
$[\varepsilon_1,\varepsilon_d]$ so that we can change variables to
\begin{align}
\mathcal{G}^{(0)}(u,v):= G(R(u),R(v))\;.
\end{align}
Comparison of (\ref{G-formula}) with 
(\ref{result-Psi}) shows, recalling $\alpha=1$,
\begin{align}
  \mathcal{G}^{(0)}(u,v)=\lim_{\epsilon\searrow 0}
  \frac{\exp(\Psi(u+\mathrm{i}\epsilon;v))}{R(v)-R(-u-\mathrm{i}\epsilon)}\;.
\end{align}
Taking Proposition \ref{prop:Psi} into acount, we have established
\begin{align}
  \mathcal{G}^{(0)}(u,v)=\frac{1}{R(u)-R(-v)}
  \prod_{k=1}^d \frac{R(v)-R(-\hat{u}^k)}{R(v)-R(\varepsilon_k)}\;.
 \label{calG-branch}
\end{align}

The representation (\ref{calG-branch}) is rational in the first variable. 
There are two ways to
proceed. First, we can expand (\ref{calG-branch}) via
(\ref{R-rational}) to
\begin{align}
\mathcal{G}^{(0)}(u,v)
\nonumber
  &= \frac{\prod_{k=1}^d (u-\varepsilon_k)}{
(u+v)\prod_{k=1}^d (u+\hat{v}^k)}
\\
& \quad \times \prod_{k=1}^d 
\frac{
(u+\hat{v}^k)\prod_{l=1}^d 
(-\hat{u}^l-\hat{v}^k)
}{
\prod_{l=1}^d (\varepsilon_l-\hat{v}^k)}
\prod_{k=1}^d \frac{\prod_{l=1}^d 
(\varepsilon_k+\varepsilon_l)}{
(u-\varepsilon_k)\prod_{l=1}^d 
(\varepsilon_k-\hat{u}^l)}
\nonumber
\\
&=\frac{1}{u+v} \prod_{k,l=1}^d \frac{(\varepsilon_k+\varepsilon_l)
(-\hat{u}^l-\hat{v}^k)}{
(\varepsilon_k-\hat{u}^l)(\varepsilon_l-\hat{v}^k)}\;.
  \label{calG-branch-1}
\end{align}
This formula is manifestly symmetric in $u,v$ --- a crucial property below.

To derive a formula which is rational in both $u,v$ we
consider the limit $u\to \varepsilon_a$ of (\ref{calG-branch}), 
which reads with $r_a=\varrho_a R'(\varepsilon_a)$:
\begin{corollary} 
\label{Cor:Gab}
For any $a=1,\dots,d$ and $v$ in a neighbourhood of $\mathbb{R}_+$ 
one has
\begin{align}
-\frac{\lambda}{{N}} r_a \mathcal{G}^{(0)}(\varepsilon_a,v)
= 
\frac{\prod_{k=1}^d (R(\varepsilon_a)-R(-\hat{v}^k))}{
\prod_{a\neq j=1}^d (R(\varepsilon_a)-R(\varepsilon_j))}\;.
\label{corGev}
\end{align}
In particular, for any $a,b=1,\dots,d$ one has
\begin{align}
\mathcal{G}^{(0)}(\varepsilon_a,\varepsilon_b)
= -\frac{{N}}{\lambda r_a}
\frac{\prod_{k=1}^d (R(\varepsilon_a)-R(-\widehat{\varepsilon_b}^k))}{
\prod_{a\neq j=1}^d (R(\varepsilon_a)-R(\varepsilon_j))}
= -\frac{{N}}{\lambda r_b}
\frac{\prod_{k=1}^d (R(\varepsilon_b)-R(-\widehat{\varepsilon_a}^k))}{
\prod_{b\neq j=1}^d (R(\varepsilon_b)-R(\varepsilon_j))}\;.
\label{corGee}
\end{align}
\end{corollary}

Next we recall the basic lemma\footnote{The rational function of $x_0$
  has potential 
simple poles at $x_0=x_k$, $k=1,\dots, d$, but all residues cancel. Hence, 
it is an entire function of $x_0$, by symmetry in all $x_k$. The
behaviour for $x_0\to \infty$ gives the assertion.}
\begin{align}
\sum_{j=0}^{d} 
\frac{\prod_{k=1}^d(x_j-c_k)}{
\prod_{j\neq k=0}^{d+1} (x_j-x_k)}=1\;,
\label{basiclemma}
\end{align}
valid for pairwise different $x_0,\dots,x_{d}$ and any
$c_1,\dots,c_d$. 
We use (\ref{basiclemma}) for $x_0=R(u)$,
$x_k=R(\varepsilon_k)$ and 
$c_k=R(-\hat{v}^k)$ to rewrite (\ref{calG-branch}) as
\begin{align}
  \label{calG-sum}
  \mathcal{G}^{(0)}(u,v) &= \frac{1}{R(v)-R(-u)} 
\Big(1+\sum_{k=1}^d \frac{1}{R(u)-R(\varepsilon_k)} 
\frac{\prod_{l=1}^d( R(\varepsilon_k)-R(-\hat{v}^l))}{
\prod_{k\neq j=1}^d (R(\varepsilon_k)-R(\varepsilon_j))}\Big)
\nonumber
\\
&= \frac{1}{R(v)-R(-u)} 
\Big(1+\frac{\lambda}{{N}}
\sum_{k=1}^d \frac{r_k \mathcal{G}^{(0)}(\varepsilon_k,v)}{
R(\varepsilon_k)-R(u)} 
\Big)\;.
\end{align}
The last line results from (\ref{corGev}).  Using the symmetry
$\mathcal{G}^{(0)}(\varepsilon_k,v)=\mathcal{G}^{(0)}(v,\varepsilon_k)$,
the previous formulae give rise to a representation of
$\mathcal{G}^{(0)}(u,v)$ which is \emph{rational in both variables}:
\begin{align}
\mathcal{G}^{(0)}(z,w) &= 
\frac{\displaystyle 
1-\frac{\lambda}{{N}}\sum_{k=1}^d \frac{r_k}{(R(\varepsilon_k)-R(-w))
(R(z)-R(\varepsilon_k))} \prod_{j=1}^d 
\frac{R(w)-R(-\widehat{\varepsilon_k}^j)}{R(w)-R(\varepsilon_j)}
}{
R(w)-R(-z)} \;.
\label{Gzw-rational}
\end{align}

\begin{proposition}
\label{prop:RFE}
The planar two-point function has the (manifestly symmetric) 
rational fraction expansion 
\begin{align}
\mathcal{G}^{(0)}(z,w)
&=\frac{1}{z+w}\bigg(1+\frac{\lambda^2}{{N}^2}
\sum_{k,l,m,n=1}^d 
\frac{C_{k,l}^{m,n}}{(z-\widehat{\varepsilon_k}^m)(w-\widehat{\varepsilon_l}^n)}
\bigg)\;,
\label{G-partialfraction}
\\
C_{k,l}^{m,n}&:= 
\frac{(\widehat{\varepsilon_k}^m +\widehat{\varepsilon_l}^n) 
r_k r_l \mathcal{G}^{(0)}(\varepsilon_k,\varepsilon_l)}{
R'(\widehat{\varepsilon_k}^m)R'(\widehat{\varepsilon_l}^n)
(R(\varepsilon_l)-R(-\widehat{\varepsilon_k}^m))
(R(\varepsilon_k)-R(-\widehat{\varepsilon_l}^n))
}\;.\nonumber
\end{align}
\begin{proof}
Expanding the first denominator in
(\ref{calG-sum}) via (\ref{R-rational}), $\mathcal{G}^{(0)}(u,v)$ has
potential poles at $u=-\hat{v}^n$ for every $n=1,\dots,d$. 
However, for $u=-\hat{v}^n$ the sum in the first line of
(\ref{calG-sum}) becomes 
$\sum_{k=1}^d \frac{1}{(R(-\hat{v}^n)-R(\varepsilon_k))}
\frac{\prod_{l=1}^d( R(\varepsilon_k)-R(-\hat{v}^l))}{
\prod_{k\neq j=1}^d (R(\varepsilon_k)-R(\varepsilon_j))}
=-1$ when using the basic lemma (\ref{basiclemma}). Consequently,
$\mathcal{G}^{(0)}(z,w)$ is regular at $z=-\hat{w}^n$ and by symmetry
at $w=-\hat{z}^n$. 

This leaves the diagonal $z+w=0$ and 
the complex lines ($z=
\widehat{\varepsilon_k}^m$, any $w$) and
($w=\widehat{\varepsilon_l}^n$, any $z$)
as the only possible poles of $\mathcal{G}^{(0)}(z,w)$.
The function $(z+w)\mathcal{G}^{(0)}(z,w)$ approaches 
$1$ for $z,w\to \infty$. Its residues at $
z=\widehat{\varepsilon_k}^m ,w=\widehat{\varepsilon_l}^n$ are
obtained from (\ref{calG-sum}):
\begin{align*}
&\Res\displaylimits_{z\to \widehat{\varepsilon_k}^m, w\to
  \widehat{\varepsilon_l}^n}
(z+w)\mathcal{G}^{(0)}(z,w)
\nonumber
\\
&= -\frac{(\widehat{\varepsilon_k}^m +\widehat{\varepsilon_l}^n)}{
(R(\varepsilon_l)-R(-\widehat{\varepsilon_k}^m))} 
\frac{\lambda r_k}{{N} R'(\widehat{\varepsilon_k}^m)}
\Res\displaylimits_{w\to  \widehat{\varepsilon_l}^n}
\mathcal{G}^{(0)}(\varepsilon_k,w)
\nonumber
\\
&=\Big(\frac{\lambda}{{N}}\Big)^2 
\frac{(\widehat{\varepsilon_k}^m +\widehat{\varepsilon_l}^n)
r_k r_l 
\mathcal{G}^{(0)}(\varepsilon_k,\varepsilon_l)
}{R'(\widehat{\varepsilon_k}^m)R'(\widehat{\varepsilon_l}^n)
(R(\varepsilon_l)-R(-\widehat{\varepsilon_k}^m))
(R(\varepsilon_k)-R(-\widehat{\varepsilon_l}^n))
} \;.
\end{align*}
The second line follows from $\mathcal{G}^{(0)}(\varepsilon_k,w)
=\mathcal{G}^{(0)}(w,\varepsilon_k)$
and (\ref{calG-sum}). 
\end{proof}
\end{proposition}

\section{Examples}

\label{sec:ex}

\subsection{A Hermitian one-matrix model}

The extreme case of a single $r_1={N}$-fold degenerate
eigenvalue $E=\frac{\mu^2}{2}\cdot \mathrm{id}$ corresponds to a 
standard Hermitian one-matrix model with measure 
$\exp(-{N}\,\mathrm{Tr}( \frac{\mu^2}{2}\Phi^2+
\frac{\lambda}{4}\Phi^4))d\Phi$. This purely quartic case 
was studied in \cite{Brezin:1977sv}. Transforming 
$M \mapsto \sqrt{{N}} \mu \Phi$ and $g=\frac{\lambda}{4\mu^4}$
brings eq.\ (3) in \cite{Brezin:1977sv} into our conventions.
The equations (\ref{rho-finite}) reduce for $E_1=\frac{\mu^2}{2}$ 
and $d=1$ to
\begin{align}
\frac{\mu^2}{2}  &=\varepsilon_1
- \frac{\lambda\varrho_1}{{N}(2\varepsilon_1)}\;, 
&
1 &= \frac{{N}}{\varrho_1} 
- \frac{\lambda\varrho_1}{{N}(2\varepsilon_1)^2}
\label{rho-degenerate}
\end{align}
with principal solution (i.e.\ $\lim_{\lambda\to 0}
\varepsilon_1=\frac{\mu^2}{2}$)
\begin{align}
\varepsilon_1&=\frac{1}{6} \big(2\mu^2+\sqrt{\mu^4+12\lambda}\big) \;,&
\varrho_1&={N}\cdot  
\frac{\mu^2\sqrt{\mu^4+12\lambda}-\mu^4+ 12\lambda}{18 \lambda}\;.
\end{align}
The other root $\widehat{\varepsilon_1}^1$ with 
$R(\widehat{\varepsilon_1}^1)
=\widehat{\varepsilon_1}^1
- \frac{\lambda\varrho_1}{{N}(\varepsilon_1
+\widehat{\varepsilon_1}^1)}=R(\varepsilon_1)=\frac{\mu^2}{2}$ 
is found to be
\begin{align}
\widehat{\varepsilon_1}^1=-\frac{1}{6}
\big(\mu^2+2\sqrt{\mu^4+12\lambda}\big)= 
\tfrac{\mu^2}{2}-2\varepsilon_1 \;.
\end{align}
The planar two-point function $G_{11}^{(0)}\equiv \mathcal{G}^{(0)}(\varepsilon_1,\varepsilon_1)$ can be evaluated via 
(\ref{corGee}) or  (\ref{calG-branch-1}) to 
\begin{align}
G_{11}^{(0)}= -\frac{1}{\lambda}\Big(\frac{\mu^2}{2}
-R(-\widehat{\varepsilon_1}^1)\Big)=
\frac{4}{3} \cdot  \frac{\mu^2+2\sqrt{\mu^4+12\lambda}}{
(\mu^2+\sqrt{\mu^4+12\lambda})^2}= 
-\frac{2\widehat{\varepsilon_1}^1}{(\varepsilon_1-
\widehat{\varepsilon_1}^1)^2}\;.
 \label{G000}
\end{align}
The result can be put into 
$G_{11}^{(0)}=\frac{1}{3\mu^2} a^2(4-a^2)$
for $a^2=\frac{2\mu^2}{\mu^2+\sqrt{\mu^4+12\lambda}}$ 
and thus agrees with the literature: 
This value for $a^2$, which
corresponds to $\frac{a^2 \lambda}{\mu^2}=\varepsilon_1-\frac{\mu^2}{2}$,
solves eq.~(17a) in 
\cite{Brezin:1977sv} for $g:=\frac{\lambda}{4\mu^4}$ 
so that (\ref{G000}) reproduces\footnote{thanks to a lucky
  coincidence: In \cite{Brezin:1977sv} expectation values of traces 
$\langle \mathrm{Tr}(M^{2p})\rangle$ are studied, whereas we consider 
$\langle M_{11} M_{11}\rangle$. For constant $E$ all moments of individual
matrix elements are equal and agree up to global rescaling by 
${N}^\delta$ with expectation values of traces.} 
eq.\ (27) in \cite{Brezin:1977sv} for $p=1$ (and the convention 
$G_{11}^{(0)}=\frac{1}{\mu^2}$ for $\lambda=0$). 

The meromorphic extension $\mathcal{G}^{(0)}(z,w)$ is most conveniently 
derived from Proposition~\ref{prop:RFE} after 
cancelling the 
two representations 
(\ref{G000}) for
$G_{11}^{(0)}=\mathcal{G}^{(0)}(\varepsilon_1,\varepsilon_1)$:
\begin{align}
\mathcal{G}^{(0)}(z,w)&=\frac{1}{z+w}\Big(1-
\frac{(\varepsilon_1+\widehat{\varepsilon_1}^1)^2}{
(z-\widehat{\varepsilon_1}^1)(w-\widehat{\varepsilon_1}^1)}\Big)
\\
&=
\frac{1}{z+w}\Big(1-\frac{\mu^4(1-a^2)^2}{
(3a^2 z+\mu^2)(3a^2 w+\mu^2)}\Big)\;,
\nonumber
\end{align}
where $a^2=\frac{2\mu^2}{\mu^2+\sqrt{\mu^4+12\lambda}}$.
We have used $R'(\widehat{\varepsilon_1}^1)=
\frac{\widehat{\varepsilon_1}^1-\varepsilon_1}{
\widehat{\varepsilon_1}^1+\varepsilon_1}$.

\subsection{A special case in $D=2$: constant 
density}

The case $\rho_0(x)\equiv \chi_{[\tilde{M}^2,\tilde{\Lambda}^2]}(x)$
was solved in \cite{Panzer:2018tvy}.
The relation (\ref{relation-rho}) shows that the measure $\varrho$
is also a characteristic function, of different support.
The dimensional classification of
Definition~\ref{def:dimension} gives $D=2$ in the limit
$\tilde{\Lambda}\to\infty$. It is convenient to adjust the free parameter $\beta$ such that (\ref{Taylor-2}) holds with 
$\varrho
=\chi_{[1,\infty)}$. Then (\ref{Taylor-2}) evaluates to
\begin{align}
R(z)= z+\lambda \log(1+z)\;.
\end{align}
The inverses are provided by the branches of Lambert-W 
\cite{Corless:1996??}, in particular
\begin{align}
R^{-1}(z)=
\lambda
W_0\Big(\frac{1}{\lambda}e^{\frac{1+z}{\lambda}}\Big)-1\;.
\end{align}
The formula (\ref{G-formula}) for $G(x,y)$ specifies to its counterpart in 
\cite{Panzer:2018tvy}.

To approach the remaining integral in $G(x,y)$ in (\ref{G-formula}) 
one could try to
approximate $R$ by a rational function. As a Stieltjes function, 
$\frac{\log(1+z)}{z}$ has uniformly convergent Pad\'e approximants
obtained by terminating the continued fraction
\[
\log(1+z)=z/(1+z/(2+z/(3+4z/(4+4z/(5+9z/(6+9z/(7+16z\dots)))))))
\]
after $2d-1$ or $2d$ fractions.

\subsection{A particular case in $D=4$: linear density} 

\label{sec:Moyal}

The case $\rho_0(x)=x\chi_{[\tilde{M}^2,\infty)}(x)$ corresponds to the
self-dual $\lambda\Phi^4$-model on four-dimensional Moyal space
\cite{Grosse:2004yu, Grosse:2012uv} and is therefore of particular
interest. The relation (\ref{relation-rho}) reads in this case
\[
  \varrho(x)  =\begin{cases}
\qquad 0 & \text{if } x<M^2\text{ or } x>\Lambda^2\;,
\\
\alpha x+\beta-\lambda \int_{M^2}^{\Lambda^2} \frac{\varrho(t) \,dt}{
  t+x} & \text{if } M^2\leq x\leq \Lambda^2\;.
\end{cases}
\]
Here an upper bound of the support (`regularisation') has been introduced, and
the lower bound $M$ depends on $\tilde{M}$ and the other parameters.
Introducing $\tilde{\varrho}_\lambda(s):=\varrho(s+M^2)$, shifting
$x\mapsto M^2+x$ and choosing
$\beta+\alpha M^2 :=\lambda \int_{0}^{\Lambda^2-M^2}dt
\frac{\tilde{\varrho}_\lambda(t)}{
  t+2M^2}$ and $\alpha=1-\lambda 
\int_{0}^{\Lambda^2-M^2}dt
\frac{\tilde{\varrho}(t)}{(t+2M^2)^2}
  $
of spectral dimension 4 one arrives at a standard Fredholm
integral equation of second kind
\begin{align}
\tilde{\varrho}_\lambda (x)
&= x - \lambda x^2 
\int_0^{\Lambda^2-M^2} \frac{\tilde{\varrho}_\lambda(s)\,ds}{
(s+2M^2)^2(s+2M^2+x)}\;.
\label{rho-Fredholm}
\end{align}
This equation is considered for $0\leq x\leq \Lambda^2-M^2$. It permits
the limit $\Lambda^2\to \infty$ corresponding  to the initial density
$\rho_0(x)=x\chi_{[\tilde{M^2},\infty)}(x)$.

In \cite{Grosse:2019qps} we prove
that (\ref{rho-Fredholm}) is for $\Lambda\to \infty$
solved by a hypergeometric function:
\begin{align}
\label{Rx-final}
\tilde{\varrho}_\lambda (x)= x  \;_2F_1\Big(\genfrac{}{}{0pt}{}{
\alpha_\lambda,\;1-\alpha_\lambda}{2}\Big|{-}\frac{x}{2M^2}\Big),
\quad
\alpha_\lambda:=\left\{ \!\!\! \begin{array}{cl} \frac{\arcsin(\lambda\pi)}{\pi}
& \text{for } |\lambda|\leq \frac{1}{\pi}\,, \\
\frac{1}{2}+\mathrm{i}  \frac{\mathrm{arcosh}(\lambda\pi)}{\pi}
& \text{for } \lambda \geq \frac{1}{\pi} \,.
\end{array} \right.
\end{align}
Remarkably, the spectral dimension $D_{spec}$ introduced in
Definition~\ref{def:dimension} gets modified by the
potential $\frac{\lambda}{4}\,\mathrm{Tr}(\Phi^4)$
from $D_{spec}(\rho_0)=4$ to
$D_{spec}(\tilde{\varrho}_\lambda)=4-\frac{2}{\pi} \arcsin(\lambda\pi)$.  For
$\lambda>0$, this dimension drop makes $R^{-1}$ globally defined on
$\mathbb{R}_+$. In this way, and in sharp contrast
\cite{Aizenman:2019yuo} to the usual $\lambda\phi^4_4$ quantum field
theory, the matricial $\lambda \Phi^{\star 4}_4$-model does not suffer
from a triviality problem.

\section{Epilogue: QFT on noncommutative spaces  and blobbed topological recursion}

\label{sec:out}

The solution of the non-linear equation for the planar 2-point function
$G_{ab}^{(0)}$ achieved in this paper is the breakthrough that now
permits a complete solution of the quartic matrix model. Many matrix models
are known to be exactly solvable, often implemented and understood in terms of
\emph{topological recursion} (TR) \cite{Eynard:2007kz}. The value of our new 
example is twofold:
\begin{enumerate}
\item It leads to a truly interacting quantum field theory in four dimensions
  \cite{Grosse:2019qps} (on a
  noncommutative space).
\item It is an example for blobbed topological recursion \cite{Borot:2015hna} in which
  the abstract loop equations \cite{Borot:2013lpa} can be proved globally
  \cite{Hock:2023nki}. 
\end{enumerate}

\subsection{The $\lambda\Phi^4$-QFT model on
  noncommutative geometry}

The statement of Theorem~\ref{thm:final} that the integral equation
(\ref{eq:Gint})  admits an exact solution (\ref{G-formula})
confirms a conjecture which crystallised during
a decade of work of two of us (HG, RW).  Building on a Ward-Takahashi
identity found in \cite{Disertori:2006nq}, we derived long ago in
\cite{Grosse:2009pa} a closed non-linear integral equation for
$G_{ab}^{(0)}$ in the large-${N}$ limit. Over the years we found so
many surprising facts about this equation that the quartic matrix
model being solvable is the only reasonable explanation.  A key step
was the reduction to an equation for an angle functions of essentially
only one variable \cite{Grosse:2012uv}.  Moreover, a recursive formula
to determine all planar $N$-point functions
$G^{(0)}_{b_0\dotsb_{N-1}}$ from the planar two-point function
$G_{ab}^{(0)}$ was found in \cite{Grosse:2012uv}.  This recursion was
later solved in terms of a combinatorial structure named `nested
Catalan table' \cite{deJong:2019oez}. In \cite{Panzer:2018tvy},
one of us (RW) with E.~Panzer obtained 
the exact solution of
$G_{ab}^{(0)}$ (at large ${N}$) in the case
$E= \mathrm{diag}(1,2,3,4,\dots)$. The solution is expressed in terms
of the Lambert function defined by the implicit equation
$W(z) \exp(W(z))=z$.

In the present paper we understood that the function
$z+\lambda\log(1+z)$ which governs the exact solution in
\cite{Panzer:2018tvy} must be generalised to a function $R$ 
which involves the 
Stieltjes transform of a \emph{deformed}
spectral measure $\varrho$, whereas for \cite{Panzer:2018tvy}
the original spectral measure $\rho_0=\chi_{[M^2,\Lambda^2]}$
was sufficient. Using classical tools such as
Cauchy's residue theorem (1831) and B\"urmann's extension
(1799) of the Lagrange inversion formula (1770) we were able to
evaluate various integrals involving $R$. The motivation to consider these
integrals comes from \cite{Panzer:2018tvy}.
It turned out that the boundary values of one of the integrals $\Psi$
provide the solution of the initial integral equation, in a striking
analogy to the Makeenko-Semenoff approach \cite{Makeenko:1991ec}
to the Kontsevich model.

\subsection{Remarks on an alternative proof for
  finite matrices}

In the case of finite matrices studied in sec.~\ref{sec:finiteN},
the function $R:\mathbb{P}^1 \to \mathbb{P}^1$
satisfies 
\begin{align}
R(z) +\frac{\lambda}{{N}}  \sum_{k=1}^d r_k
\mathcal{G}^{(0)}(z,\varepsilon_k)
+\frac{\lambda}{{N}}  \sum_{k=1}^d
\frac{r_k}{R(\varepsilon_k)-R(z)}
=-R(-z)\;.
\label{Rzz}
\end{align}
Indeed, the original equation (\ref{Gab-orig}) 
for $G^{(0)}_{ab}
=\mathcal{G}^{(0)}(\varepsilon_a,\varepsilon_b)$ with 
$\mathcal{O}({N}^{-1})$-contributions dropped (in accordance 
with planarity) extends to 
complex variables $\varepsilon_a \mapsto z$ and 
$\varepsilon_b\mapsto w$:
\begin{align}
  \label{Gzw-orig}
  &\Big\{R(z)+R(w) +\frac{\lambda}{{N}}  \sum_{k=1}^d r_k
\mathcal{G}^{(0)}(z,\varepsilon_k)
+\frac{\lambda}{{N}}  \sum_{k=1}^d
\frac{r_k}{R(\varepsilon_k)-R(z)}\Big\} \mathcal{G}^{(0)}(z,w)
\\[-1ex]
&= 1+ 
\frac{\lambda}{{N}}  \sum_{k=1}^d
\frac{r_k\,\mathcal{G}^{(0)}(\varepsilon_k,w)}{
R(\varepsilon_k)-R(z)} \;.
\nonumber
\end{align}
Now (\ref{Rzz}) follows by comparison with (\ref{calG-sum}). 
Equation (\ref{Rzz}), with $\mathcal{G}^{(0)}$ scaled by $\alpha^{-1}$, has 
also been established for
H\"older-continuous measure in Theorem~\ref{thm:R} together
with (\ref{Psi-G}).

In \cite{Schurmann:2019mzu} the converse approach is pursued. It is
\emph{supposed} that there exists a rational function $R$ which
satisfies (\ref{Rzz}) plus some technical assumptions.  Then
(\ref{calG-branch}) and the equation (\ref{Rx-N}) for $R$ is deduced,
and finally the consistency of the ansatz (\ref{Rzz}) is shown. In this way
(\ref{Gzw-rational}) and the structure (\ref{Rx-N}) of $R$
are directly proved without consideration of
boundary value problems. Of course, one would never have guessed
the ansatz  (\ref{Rzz}) without the insight from the present paper.

\subsection{Blobbed topological recursion}

The solution (\ref{Gzw-rational}) of the planar 2-point function,
combined with previous work
\cite{Grosse:2012uv, deJong:2019oez}, shows that all planar moments
(\ref{moments-P}), i.e.\ of topology of a disc 
$(g=0, n=1)$,
can be exactly
solved, as \emph{convergent} functions of $\lambda$, for any operator $E$ (of
spectral dimension $\leq 4$). After simplifications in
\cite{Schurmann:2019mzu} (and solution of the planar 1+1-point
function), it was understood in \cite{Branahl:2020yru} that the
solution of the (after $1/N$-expansion) affine equations for all other
moments (\ref{moments-P}) needs and defines a family $\Omega^{(g)}_n$
of auxiliary functions which together with two other auxiliary families
satisfies a coupled system of equations. The solution for small $g+2n$
suggested that the meromorphic differentials
$\omega_{g,n}(z_1,...,z_n)=\Omega^{(g)}_n(z_1,...,z_n) dR(z_1)\cdots
dR(z_n)$ obey blobbed topological recursion (BTR), an extension of
topological recursion due to Borot and Shadrin \cite{Borot:2015hna}.
The initial data $(x,y,\omega_{0,2})$ of the spectral curve are
$x(z)=R(z)$, $y(z)=-R(-z)$ and
$\omega_{0,2}(z_1,z_2)=\frac{dz_1\,dz_2}{(z_1-z_2)^2}+
\frac{dz_1\,dz_2}{(z_1+z_2)^2}$.  The conjecture that the quartic
analogue of the Kontsevich model satisfies BTR has been proved by two of us (AH+RW)
for $g=0$ in \cite{Hock:2021tbl} and (with techniques inspired from
the Hermitian 2-matrix model \cite{Chekhov:2006vd}) for $g=1$ in
\cite{Hock:2023nki}.  In particular, linear and and quadratic loop
equations \cite{Borot:2013lpa} for the $\omega_{g,n}$ have been
established globally on the Riemann sphere.  It is a general fact
\cite{Borot:2015hna} that
$\omega_{g,n}$ satisfying BTR encode intersection numbers on
$\overline{\mathcal{M}}_{g,n}$.
A first link
to the BKP integrable hierarchy was found in \cite{Borot:2023thu}.

  The Langmann-Szabo-Zarembo model \cite{Langmann:2003if} is a variant
  with complex (instead of Hermitian) matrices of the matricial
  QFT-model considered here.  In \cite{Branahl:2022uge} it is shown
  that that the LSZ model leads to a variant of (\ref{Rzz}), which is
  solved by similar techniques. Then a family $\omega_{g,n}$ of
  meromorphic differentials is obtained which is proved to follow
  standard topological recursion.

\subsection{Implications for QFT in 4 dimensions}
  
\label{sec:implications}

A main challenge in QFT is to
construct an interacting model in 4 dimensions. Aizenman and
Duminil-Copin recently proved \cite{Aizenman:2019yuo} that the rather
simple $\lambda\phi^4_4$-model is \emph{not} a valid example: it is
marginally trivial, hence non-interacting in the limit to continuum
and infinite volume. It is expected that non-Abelian Yang-Mills theory
will provide a valid example, but the proof of this conjecture is one
of the millenium prize problems. Euclidean quantum field theories on
noncommutative geometries \cite{SurveyNCG} provide a new class of
examples to try. They
violate the axioms related to Euclidean invariance, but their
behaviour under renormalisation is very close to traditional QFT.
In fact the situation is better: In a subsequent work
\cite{Grosse:2019qps} we showed that in the $\lambda\Phi^4_4$-model on
noncommutative Moyal space (at large deformation), the solution
(\ref{Rx-final}) of the deformation equation (\ref{relation-rho}) 
implies a reduction of the effective spectral dimension  from the na\"ive
value $4$ to
$4-\frac{2}{\pi}\arcsin(\lambda \pi)$. As consequence of the dimension drop,
this model defines a
non-trivial (i.e.\ truly interacting) just-renormalisable QFT in 4
dimensions (on a quantum space, though). It would be interesting
to investigate whether the reduced spectral dimension, consequence of our
exact solution of the two-point function, permits to transfer the spectacular
methods and results \cite{Hairer2014, Mourrat:2016vlt, Gubinelli:2021nou}
of the ordinary $\lambda\phi^4_3$-model to
the 4-dimensional noncommutative case.


\newcommand{\etalchar}[1]{$^{#1}$}

\end{document}